\documentclass{article}
\usepackage{arxiv}

\usepackage[utf8]{inputenc} 
\usepackage[T1]{fontenc}    
\usepackage{hyperref}       
\usepackage{url}            
\usepackage{booktabs}       
\usepackage{amsfonts}       
\usepackage{nicefrac}       
\usepackage{microtype}      
\usepackage{lipsum}		
\usepackage{graphicx}
\usepackage{natbib}
\usepackage{doi}

\DeclareMathAlphabet\mathsfbi            {T1}{phv}{b}{it}
\usepackage{epstopdf, epsfig}
\usepackage{amsmath}

\DeclareMathOperator{\csch}{csch}

\newcommand\Real{\mbox{Re}}          
  
\newcommand{\R}{\mathbb{R}}
\newcommand{\C}{\mathbb{C}}
\newcommand{\N}{\mathbb{N}}
\newcommand{\Z}{\mathbb{Z}}

\author{ \href{https://orcid.org/0000-0003-4916-8392}{\includegraphics[scale=0.06]{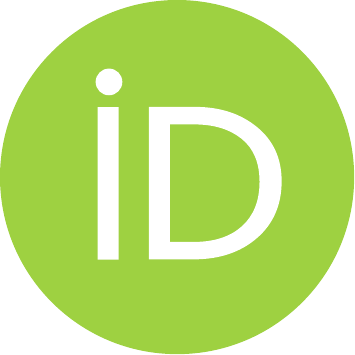}\hspace{1mm}Ben Wilks} \\
Department of Mathematics and Statistics\\
University of Otago\\
PO Box 56, Dunedin 9054, New Zealand\\
\texttt{wilbe612@student.otago.ac.nz} \\
	\And
	\href{https://orcid.org/0000-0002-6126-6385} {\includegraphics[scale=0.06]{orcid.pdf}\hspace{1mm}Fabien Montiel} \\
Department of Mathematics and Statistics\\
University of Otago\\
PO Box 56, Dunedin 9054, New Zealand\\
\texttt{fmontiel@maths.otago.ac.nz} \\
	\AND
	\href{https://orcid.org/0000-0002-5967-4267}{\includegraphics[scale=0.06]{orcid.pdf}\hspace{1mm}Sarah Wakes} \\
Department of Mathematics and Statistics\\
University of Otago\\
PO Box 56, Dunedin 9054, New Zealand\\
\texttt{sarah.wakes@otago.ac.nz } \\
}

\date{}

\title{Graded arrays of vertical barriers: rainbow reflection and broadband energy absorption}

\begin{document}

\maketitle

\begin{abstract}
The rainbow reflection effect describes the broadband spatial separation
of wave spectral components caused by a spatially graded array of
resonators. Although mainly studied in optics and acoustics, this
phenomenon has recently been demonstrated both theoretically and
experimentally for water waves travelling through an array of vertical
cylinders. Linear water wave scattering by a array of vertical,
surface-piercing barriers is considered here, in which both the
submergence and spacing between the barriers are spatially graded. The
rainbow reflection effect arises naturally as wave energy temporarily
becomes amplified at different locations depending on frequency. Band
diagram calculations are used to demonstrate that this is a consequence
of the wave gradually slowing down throughout the array. The
wave/barriers scattering problem is then augmented by positioning
heave-restricted, rectangular floating bodies equipped with a linear
damping mechanism between each adjacent pair of barrier. A solution to
the resulting boundary-value problem is obtained using an integral
equation/Galerkin method. Using constrained optimisation, passive
rainbow absorbers are designed that achieve near-perfect absorption over
(i) a discrete set of frequencies, and (ii) over an octave. This
suggests potential applications of rainbow absorbers in the design of
smart coastal technologies.
\end{abstract}

\section{Introduction}
There is growing interest throughout the wave sciences in metamaterials which have a subwavelength structure that is spatially varying in some way. This spatial variation can give rise to so called rainbow phenomena, through which broadband wave signals are slowed and spatially separated into their spectral components. These phenomena can be classified into rainbow trapping, where the separated energy becomes permanently localised, and rainbow reflection, where the energy is temporarily amplified before being reflected \citep{He2012,Chaplain2020}. First demonstrated by \citet{Tsakmakidis2007} for optical waves, these concepts have since inspired the development of analogues in several fields, including elastic waves \citep{Skelton2018,Arreola-Lucas2019}, seismic waves \citep{Colombi2016}, water waves \citep{Bennetts2018} and acoustics. In this latter discipline, rainbow reflection has been prominent in the design of acoustic filters and sensors, where one-dimensional waveguides with graded resonant side-cavities have been shown to spatially separate frequency components \citep{Zhu2013,Ni2014,Zhao2019}. The local energy amplification of acoustic waves has also been studied in two-dimensional arrays of (i) rigid cylinders with chirped spacing \citep{Romero-Garcia2013,Cebrecos2014} and (ii) C-shaped resonators with graded radii \citep{Bennetts2019}. \citet{Jimenez2017} further extended the study of rainbow reflecting devices by including a loss-inducing mechanism within each resonator of a graded array. Using numerical optimisation, they created a structure that achieves near-perfect sound absorption over a prescribed frequency interval. Our primary goal is to investigate the feasibility of achieving this rainbow absorption for water waves.

Recently, several other remarkable capabilities of metamaterials, originally studied for optical or acoustic waves, have been predicted in the context of water waves, where research is motivated by applications that include the development of coastal protection technology or wave energy parks. For instance, two-dimensional periodic arrays of uniform rigid cylinders immersed in water exhibit negative refraction, ultra-refraction and bandgaps---the latter characterising frequency intervals in which waves cannot propagate without a change in amplitude \citep{McIver2000,Hu2004,Farhat2010}. The lowest frequency bandgap of such arrays is shifted to lower frequencies when the cylinders are replaced with C-shaped cylindrical resonators, which suggests that they could be used as effective breakwaters \citep{Hu2011,Dupont2017}. \citet{Bennetts2018} demonstrated rainbow reflection in arrays of C-shaped cylindrical resonators with graded radii. In these arrays, the group velocity of the waves gradually approaches zero through the array at a location that depends on frequency, which results in local energy amplification. Arrays of rigid cylinders with chirped spacing also give rise to the rainbow reflection effect and associated local energy amplifications, which was verified experimentally by \citet{Archer2020}. Here, the absence of a resonant substructure means that this effect is limited to higher frequency intervals when compared with arrays of C-shaped cylinders. Although local energy amplification in graded and periodic arrays is now fairly well understood, energy absorption within such systems remains mostly unexplored.

In the last few decades, wave energy converters have been the subject of an intensive research effort \citep{Falcao2010}. Several devices have been proposed which are capable of absorbing 100\% of wave energy at a particular frequency, but their efficiency is usually sensitive to a small change in frequency \citep{Salter1974,Evans1979,Evans1995}. Because real ocean waves occur over a broad range of frequencies, the limited band of effectiveness of single absorbers is typically overcome using optimal control theory \citetext{see \citealp{Coe2017} and references therein}. Recent research has also focused on the design of arrays of multiple absorbers, or wave energy parks, which optimally extract energy from realistic sea states \citep{Babarit2013,Goteman2020}. By using shallow water theory, \citet{Porter2021} recently demonstrated perfect absorption at all frequencies by a semi-infinite array of damped buoys. The buoys were modelled using a homogenised boundary condition at the upper surface of the two-dimensional fluid domain. To achieve perfect absorption, the first segment of the absorbing surface must have a negative damping coefficient and thus add energy to the fluid. Using full linear water wave theory, a similar device of finite length was shown to achieve near perfect absorption for sufficiently large frequencies.

In \textsection\ref{graded_arrays}, we explore rainbow reflection in a one-dimensional waveguide consisting of a graded array of parallel, surface-piercing rigid vertical barriers. The limited complexity of this problem compared with two-dimensional arrays allows us to reveal a clear relationship between the structural geometry and the local energy amplifications. We build upon the results of \citet{Ursell1947} and \citet{Porter1995a}, who studied the scattering of water waves by a single vertical barrier, and those of \citet{Evans1972}, \citet{Newman1974} and \citet{McIver1985}, who studied resonators consisting of two vertical barriers. In the present study, we demonstrate how these resonators can be combined into an array, and how local energy amplification within this array relates back to the local geometry, as well as to the band structure of the corresponding periodic infinite array.

The local energy amplification in graded arrays of vertical barriers suggests that highly efficient broadband absorption could be achieved by positioning suitable energy absorbing mechanisms within the array. In \textsection3 we modify the problem so that the surface between each successive pair of barriers is occupied by a damped rectangular piston that is restricted to move in heave. Our solution to the corresponding linear boundary value problem is obtained by reformulating it as a coupled system of integral equations, which we solve using a Galerkin method. In \textsection4, we briefly discuss absorption by a single piston oscillating between two barriers. Then, we design rainbow absorbers which achieve near-perfect absorption over (i) a discrete set of frequencies, and (ii) over an octave. Our results extend those from acoustics obtained by \citet{Jimenez2017}, and highlight the importance of non-absorbing structures in transferring and localising energy.

\section{Graded arrays of vertical barriers}\label{graded_arrays}
\subsection{Preliminaries}
We consider the scattering of linear water waves by $N+1$ parallel surface-piercing vertical barriers in a fluid domain of infinite horizontal extent and constant, finite depth $H$. A two-dimensional Cartesian coordinate system $(x,z)$ is used, where the $z$-axis points vertically upwards and the horizontal coordinate $x$ corresponds to the direction of propagation of the waves. The fluid domain is bounded vertically by a flat sea-bed at $z=-H$ and a free surface located at $z=0$ when at equilibrium.

\begin{figure}
  \centerline{\includegraphics[width=\textwidth]{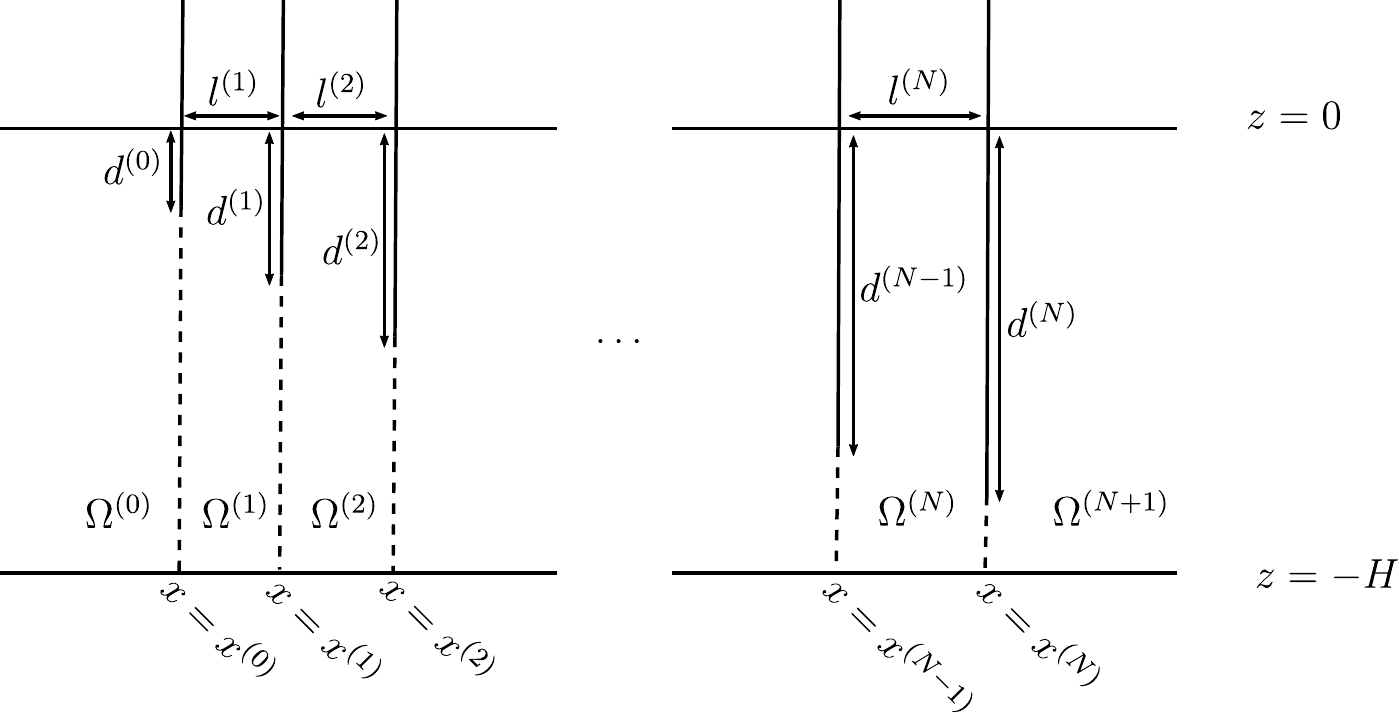}}
  \caption{Schematic of the multiple vertical barrier problem.}
\label{fig:schematic1}
\end{figure}

Each vertical barrier is of infinitesimal thickness and is situated at $x=x^{(n)}$, where $x^{(n)}\in\R$ for all $n\in\{0,\dots,N\}$, and $x^{(n)}>x^{(n-1)}$ for $n\geq 1$. Each barrier is defined by the set of points $\Gamma_b^{(n)}=\{(x^{(n)},z)|z\geq -d^{(n)}\}$, where $d^{(n)}\in(-H,0)$ is the submergence depth of the $n$\textsuperscript{th} barrier. A schematic of this geometry is given in figure \ref{fig:schematic1}.

We denote by $\Omega$ the region occupied by the fluid at equilibrium, that is
\begin{equation}
    \Omega=\left\{(x,z)\in\R\times(-H,0)\left|(x,z)\notin\bigcup_{n=0}^{N}\Gamma_b^{(n)} \right.\right\}.
\end{equation}
We also define the following subregions
\begin{align*}
    \Omega^{(0)} &= \{(x,z)|x<x^{(0)}\text{ and }z\in(-H,0)\},&\\
    \Omega^{(n)} &= \{(x,z)|x\in(-x^{(n-1)},x^{(n)})\text{ and }z\in(-H,0)\}&\text{for all }n\in\{1,\dots,N\},\\
    \Omega^{(N+1)} &= \{(x,z)|x>x^{(N)}\text{ and }z\in(-H,0)\}.&
\end{align*}

In order to apply linear water wave theory to this problem, we assume that the steepness of the free surface elevation of the incident waves is small. Thus, we consider the fluid to be inviscid, incompressible and undergoing irrotational, time-harmonic motion with angular frequency $\omega$. We assume that there is no normal flow across the seabed and the barriers. The velocity field of the fluid can therefore be expressed as the spatial gradient of some potential $\Phi:\Omega\times\R\to\R$, which is of the form
\begin{equation}
    \Phi(x,z,t) = \Real\left(\phi(x,z)\exp(-\mathrm{i}\omega t)\right).
\end{equation}
The complex-valued spatial function $\phi$ then satisfies the following boundary value problem
\begin{subequations}
\begin{align}
    \bigtriangleup\phi&=0 &\forall(x,z)\in \Omega\label{laplacian}\\
    \partial_z\phi&=0 & \text{for }z=-H\label{seabed}\\
    \partial_z\phi&=\frac{\omega^2}{g}\phi & \text{for } z=0\label{free_surface}\\
    \partial_x\phi&= 0 & \text{for all }(x,z)\in\bigcup_{n=0}^N \Gamma_b^{(n)}, \label{vanishing_barrier}
\end{align}
\end{subequations}
where \eqref{free_surface} is the linear free-surface condition with $g$ denoting acceleration due to gravity.

Since $\phi$ is twice differentiable, it must be continuous and have a continuous first partial derivative with respect to $x$ across the gaps beneath each barrier. This can be expressed in terms of one-sided limits as
\refstepcounter{equation}
$$
    \lim_{x\to x^{(n)-}}\phi(x,z)=\lim_{x\to x^{(n)+}}\phi(x,z)
    \quad\mbox{and}\quad
    \lim_{x\to x^{(n)-}}\partial_x\phi(x,z)=\lim_{x \to x^{(n)+}}\partial_x\phi(x,z),
    \eqno{(\theequation{\mathit{a},\mathit{b}})}\label{limits}
$$
for all $n=0,\dots,N$ where $z\in(-H,-d^{(n)})$. We require that $|\phi|$ remains finite as ${x\to\pm\infty}$. The solution must also obey an appropriate Sommerfeld radiation condition, which enforces that the scattered plane waves must propagate away from the array of barriers (see \eqref{sommerfeld}).

The submerged edges of the barriers located at $(x^{(n)},-d)$ create singularities of order $-1/2$ in the fluid velocity \citep{Evans1972,Mosig2018}. Thus we require that for all $n\in\{0,\dots,N\}$
\begin{equation}\label{corner_singularity}
    \left((x-x^{(n)})^2+(z+d^{(n)})^2\right)^{1/2}\nabla \phi\to \boldsymbol{0}\quad\mbox{\ as\ }\left((x-x^{(n)})^2+(z+d^{(n)})^2\right)^{1/2}\to 0.
\end{equation}

The general solution to the \eqref{laplacian}--\eqref{free_surface} is obtained using separation of variables
\begin{equation}\label{separation_sol1}
    \phi(x,z) = \begin{cases}
    \begin{aligned}
    \sum_{m=0}^\infty& (A_m^{(n)}\exp(\mathrm{i}k_m(x-x^{(n)}))\\
    &+B_m^{(n)}\exp(-\mathrm{i}k_m(x-x^{(n)})))\psi_m^{(0)}(z)
    \end{aligned}& \begin{aligned} \mbox{for\ } (x,z)\in\Omega^{(n)}\\
    \mbox{where }n\in\{0,\dots,N\}\end{aligned}\\
    &\\
    \begin{aligned}
    \sum_{m=0}^\infty& (A_m^{(N+1)}\exp(\mathrm{i}k_m(x-x^{(N)}))\\
    &+B_m^{(N+1)}\exp(-\mathrm{i}k_m(x-x^{(N)})))\psi_m^{(0)}(z)
    \end{aligned}& \mbox{for\ } (x,z)\in\Omega^{(N+1)}.
    \end{cases}
\end{equation}
The quantities $k_m$ are the solutions to the free surface dispersion relation ${k\tanh kH=\omega^2/g}$. We use the convention that ${k_0=2\pi/\lambda\in\R_+}$, where $\lambda$ is the wavelength of the propagating waves, and that ${-\mathrm{i}k_n\in((n-1)\pi/H,n\pi/H)}$ for all ${n\in\N}$. The vertical eigenfunctions are defined as
\begin{equation}
    \psi_m^{(0)}(z)=\beta_m^{-1/2}\cosh(k_m(z+H)),
\end{equation}
where the superscript $(0)$ has been included for consistency with \textsection\ref{solution_multiple_pistons}. The normalisation constants $\beta_m=\sinh(2k_m H)/(4k_m H)+\tfrac{1}{2}$ are chosen so that the vertical eigenfunctions satisfy
\begin{equation}\label{orthonormality}
    \int_{-H}^0\psi_m^{(0)}(\xi)\psi_p^{(0)}(\xi)d\xi = H\delta_{m p},
\end{equation}
for all $m,p\in\{0\}\cup\N$, where $\delta_{mp}$ is the Kronecker delta.

The fluid motion is forced by incident plane waves with angular frequency $\omega$. This forcing determines the incident potential on the exterior regions of the fluid, which is of the form
\begin{equation}\label{incident_field}
    \phi_{\textrm{in}}(x,z)= \begin{cases} A_0^{(0)}e^{\mathrm{i}k_0(x-x^{(0)})}\psi_0^{(0)}(z)&\mbox{\ for\ }(x,z)\in\Omega^{(0)}\\
    B_0^{(N+1)} e^{-\mathrm{i}k_0(x-x^{(N)})}\psi_0^{(0)}(z)&\mbox{\ for\ }(x,z)\in\Omega^{(N+1)}.
    \end{cases}
\end{equation}
We will mostly consider problems where the forcing originates from the left. In this case, $B_0^{(N+1)}=0$ and the transmission and reflection coefficients are $T=A_0^{(N+1)}/A_0^{(0)}$ and $R=B_0^{(0)}/A_0^{(0)}$, respectively. Since $|\phi|$ remains finite as $x\to\pm\infty$, we must have $A_m^{(0)}=B_m^{(N+1)}=0$ for all $m\geq 1$. The Sommerfeld radiation condition is thus
\begin{equation}\label{sommerfeld}
    \phi(x,z)-\phi_{\textrm{in}}(x,z)\sim \begin{cases} B_0^{(0)}e^{-\mathrm{i}k_0(x-x^{(0)})}\psi_0^{(0)}(z)&\mbox{\ as\ }x\to-\infty\\
    A_0^{(N+1)}e^{\mathrm{i}k_0(x-x^{(N)})}\psi_0^{(0)}(z)&\mbox{\ as\ }x\to\infty,
    \end{cases}
\end{equation}
for all $z\in(-H,0)$.
\subsection{Overview of the solution technique}\label{solution_overview}
We invoke the continuity conditions (\ref{limits}a,b) and the no flow condition \eqref{vanishing_barrier} for each barrier in order to map the amplitudes of waves incident on the $n$th barrier to the amplitudes of the waves scattered by that barrier. This mapping is approximated by the scattering matrix relation
\begin{equation*}
    \mathsfbi{S}^{(n,n+1)}\begin{bmatrix} \boldsymbol{A}^{(n)}\\\boldsymbol{B}^{(n+1)}\end{bmatrix}=\begin{bmatrix} \boldsymbol{A}^{(n+1)}\\\boldsymbol{B}^{(n)}\end{bmatrix}
\end{equation*}
for $n\in\{0,\dots,N\}$. Here, $\boldsymbol{A}^{(n)}$ and $\boldsymbol{B}^{(n)}$ are vectors with entries $A_m^{(n)}$ and $B_m^{(n)}$, for $0\leq m \leq M$, where $M$ is a truncation parameter.

The entries of each scattering matrix $\boldsymbol{S}^{(n,n+1)}$ are determined as a function of the barrier submergence $d^{(n)}$ as follows: (i) express the scattered wave amplitudes in terms of the horizontal fluid velocity $u(z)=\partial_x\phi$ on the interval $(-H,-d^{(n)})$, (ii) invoke the continuity conditions to construct an integral equation satisfied by $u$, and (iii) solve the integral equation via a Galerkin method which involves expanding $u$ over a basis of weighted Chebyshev polynomials, which encapsulate the singularities at $(x^{(n)},-d^{(n)})$ \citetext{see \citealp{Porter1995a} for more details}.

The scattering matrices $\{\boldsymbol{S}^{(0,1)},\dots,\boldsymbol{S}^{(N,N+1)}\}$ and the horizontal positions of the barriers $\{x^{(0)},\dots,x^{(N)}\}$ are used to assemble the global scattering matrix for the entire array of barriers, which satisfies
\begin{equation*}
    \mathsfbi{S}^{(0,N+1)}\begin{bmatrix} \boldsymbol{A}^{(0)}\\\boldsymbol{B}^{(N+1)}\end{bmatrix}=\begin{bmatrix} \boldsymbol{A}^{(N+1)}\\\boldsymbol{B}^{(0)}\end{bmatrix}.
\end{equation*}
The global scattering matrix is used to compute the amplitudes of the scattered waves in $\Omega^{(0)}$ and $\Omega^{(N+1)}$. Finally, the wave amplitudes in the interior regions $\Omega^{(n)}$ are computed recursively for $n\in\{1,\dots,N\}$. The algorithm used to construct the global scattering matrix and compute the unknown coefficients is based on the method developed by \citet{Ko1988} to study electromagnetic wave propagation in layered media. This method has been adapted for problems involving the multiple scattering of linear water waves, yielding numerically stable results \citep{Bennetts2009,Montiel2015}.

\subsection{The fundamental resonance of a pair of barriers}\label{fund_res_sec}
\begin{figure}
    \centering
    \includegraphics[width=\textwidth]{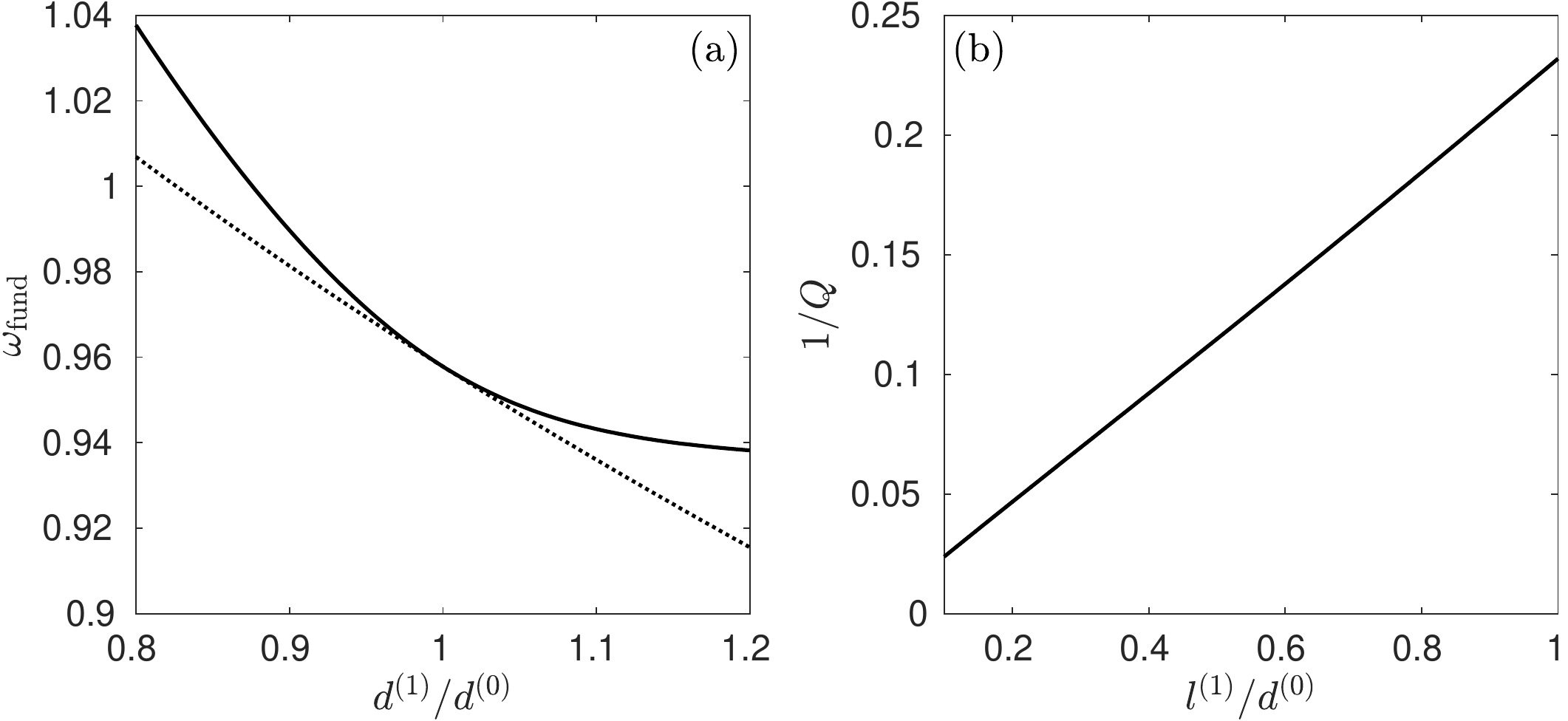}
    \caption{(a) A comparison of the fundamental frequency of two unequal barriers (solid curve) with the fundamental frequency of two equal barriers with submergence $(d^{(0)}+d^{(1)})/2$ (dotted curve). Here, $l^{(1)}=1$ is fixed. (b) A plot of $1/Q$ as a function of $l^{(1)}/d^{(0)}$. There is a very strong agreement between the data used to generate this plot and the model $1/Q=a\left(l^{(1)}/d^{(0)}\right)$, where $a\in\R$ is a coefficient. This demonstrates that $Q$ and $l^{(1)}/d^{(0)}$ are inversely proportional. In both subfigures we use $d^{(0)}=10$\,m, and the depth of the fluid is fixed to be $H=20$\,m here and throughout this paper.}
    \label{fig:lQ}
\end{figure}

We first consider the response of an isolated resonator consisting of a pair of barriers, i.e. when $N=1$. This problem has been extensively studied when $d^{(0)}=d^{(1)}$. Such a pair of identical barriers is associated with an infinite collection of resonant frequencies \citep{Evans1972}. Here, we are exclusively concerned with the fundamental resonance (i.e. the lowest frequency resonance), which is analogous to the Helmholtz resonance in acoustic resonators. The corresponding fundamental frequency, which we denote $\omega_{\textrm{fund}}$, can be approximated by $\sqrt{g/d^{(0)}}$. This approximation is valid when the barriers are close to each other and submerged in deep water, i.e. $k_0 l^{(1)}<<1$ and $d^{(0)}/H<<1$. Under these assumptions, the mass of fluid between the barriers can be treated as a heaving rectangular body oscillating about its hydrostatic equilibrium \citep{Newman1974}. As $l^{(1)}$ increases, this approximation loses validity and the fundamental frequency decreases \citep{McIver1985}. When $d^{(0)}\neq d^{(1)}$, the fundamental frequency of the corresponding resonator is similar to that of a pair of identical barriers with submergence $(d^{(0)}+d^{(1)})/2$, provided $|d^{(1)}/d^{(0)}-1|<<1$. This result is shown in figure \ref{fig:lQ}a. Throughout this paper, we fix the depth of the fluid is to be $H=20$\,m.

The bandwidth of the fundamental resonance of a pair of identical barriers is governed by the ratio $l^{(1)}/d^{(0)}$. To quantify this relationship, we first introduce the following analogue for the average power of the oscillations between two barriers
\begin{equation}
    \Tilde{P}(\omega)=\omega^2\int_{x^{(0)}}^{x^{(1)}}|\phi(x,0)|dx,
\end{equation}
where $\phi$ is the velocity potential of the fluid when excited by a plane wave of amplitude $0.1$\,m incident from the left. We define $\Delta\omega_{\textrm{fund}}$ to be the full width at half the maximum of the peak of $\Tilde{P}$ associated with the fundamental resonance. The $Q$-factor is then $Q =\omega_{\textrm{fund}} /\Delta\omega_{\textrm{fund}}$. As shown in \ref{fig:lQ}b, $Q$ and $l^{(1)}/d^{(0)}$ are inversely proportional. This extends the assertion of \citet{McIver1985} that the fundamental resonance `detunes' as the spacing between the barriers increases, which was quantified by the distance between the zeros of the reflection and transmission coefficients associated with the resonance. To the best of our knowledge, the inverse relationship between $Q$ and $l^{(1)}/d^{(0)}$ has not been reported before.

\subsection{Local energy amplifications in graded arrays}\label{local_energy_amps_sec}
The observations from the previous section motivate the design of a rainbow trapping device consisting of multiple vertical barriers, where the submergence and spacing of the barriers increases throughout the array. We aim to adjust the parameters $d^{(n)}$ and $l^{(n)}$ so that the fundamental resonances of each cavity are evenly spaced over the octave $\omega\in[0.8,1.6]$\,s$^{-1}$. The resonant frequencies of the cavities must decrease in the direction of wave propagation, in order to prevent the deep barriers of the low frequency cavities from sheltering the high frequency cavities. This is because the transmission coefficient of the $n$th barrier in isolation is a decreasing function of $\omega^2d^{(n)}$ \citep{Ursell1947}. We design the array so as to excite local energy amplifications when the waves are incident from the left. Thus for $N=9$ the fundamental resonance of the $n$th cavity is given by $\omega^{(n)}=1.6-0.1(n-1)$. 

First, we find the submergence $\tilde{d}^{(n)}$ for pairs of identical barriers so that the fundamental resonance is $\omega^{(n)}$, while fixing the separation of the barriers to be $\tilde{l}^{(n)}=\tilde{d}^{(n)}/5$. This is done by maximising $\Tilde{P}(\omega^{(n)})$ as a function of $\tilde{d}^{(n)}\in(0,H)$, which yields the optimised submergences of the barriers presented in table \ref{tab:sub_depth}. Second, we use these optima to choose parameters for the array of barriers. We want the geometry of the $n$th cavity to be similar to the pair of identical barriers which produces a resonance at $\omega^{(n)}$. Thus, we select $l^{(n)}=\tilde{l}^{(n)}$. In addition, we require that the average submergence of the two barriers forming each cavity is equal to the depth of the optimised pair of barriers, that is
\begin{equation}\label{underdetermined_sys}
    \frac{d^{(n-1)}+d^{(n)}}{2}=\tilde{d}^{(n)}
\end{equation}
for all $n\in\{1,\dots,N\}$. This is an underdetermined system for the $N+1$ unknown parameters $d^{(n)}$. We choose the solution which satisfies the constraints ${0<d^{(1)}<\dots<d^{(N)}<H}$ and minimises the grading of the array (see appendix \ref{appA} for further details).

\begin{table}
  \begin{center}
\def~{\hphantom{0}}
\begin{tabular}{cccccccccc}
$n$                       & 1    & 2    & 3    & 4    & 5    & 6    & 7    & 8     & 9     \\
$\omega^{(n)}$ (s$^{-1}$) & 1.6  & 1.5  & 1.4  & 1.3  & 1.2  & 1.1  & 1.0  & 0.9   & 0.8   \\
$g/(\omega^{(n)})^2$ (m)  & 3.83 & 4.36 & 5.01 & 5.80 & 6.81 & 8.11 & 9.81 & 12.11 & 15.33 \\
$\tilde{d}^{(n)}$ (m)     & 3.44 & 3.90 & 4.47 & 5.18 & 6.07 & 7.22 & 8.74 & 10.78 & 13.58
\end{tabular}
  \caption{The submergences $\tilde{d}^{(n)}$ of pairs of identical barriers whose fundamental resonance occurs at $\omega^{(n)}$, which were obtained by maximising the time-average power of free surface oscillations. The optimised submergence values occur at approximately 0.9 times the estimate $g/(\omega^{(n)})^2$.}
  \label{tab:sub_depth}
  \end{center}
\end{table}

\begin{figure}
  \centerline{\includegraphics[width=\textwidth]{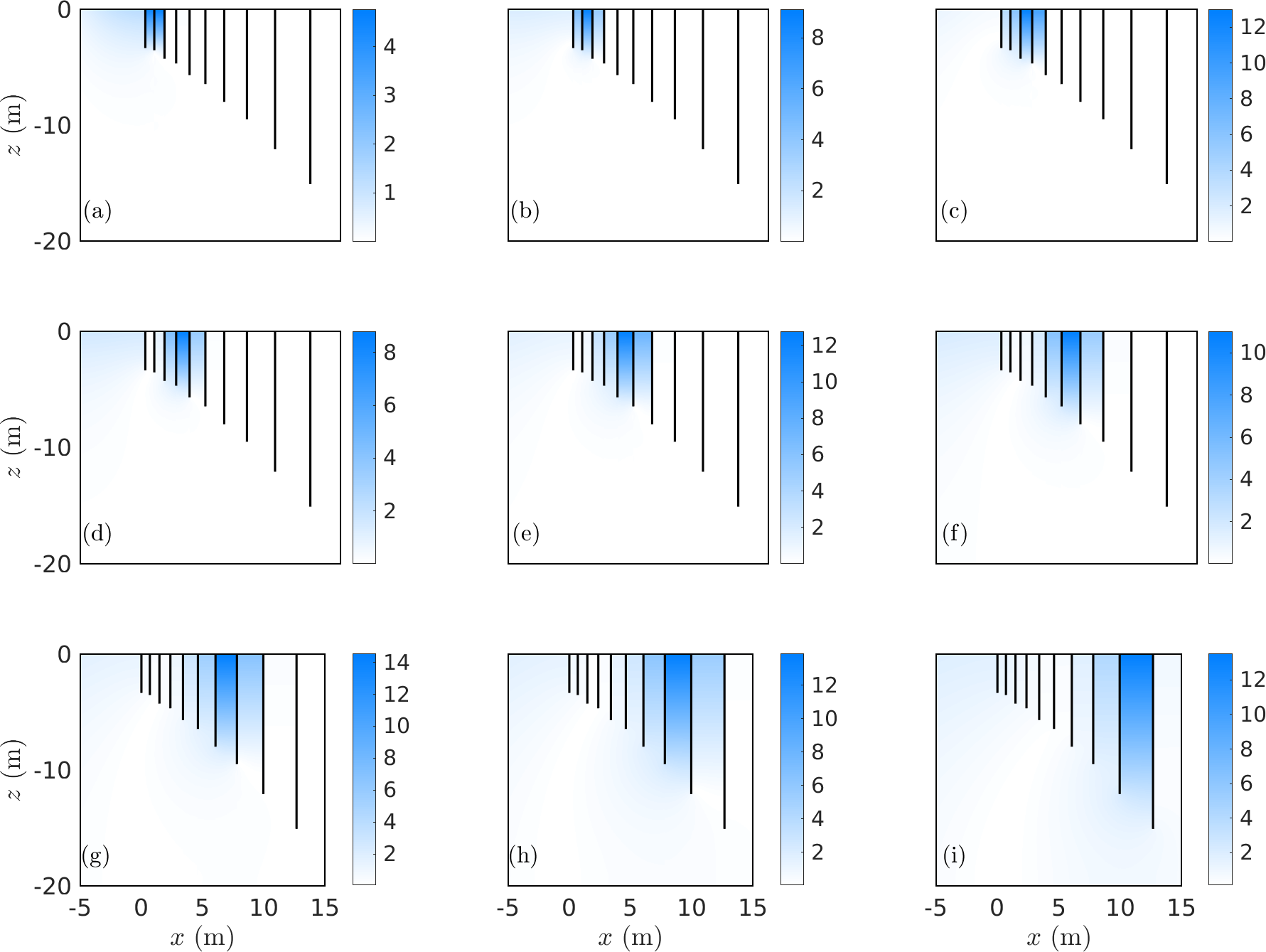}}
  \caption{Colour plots of the velocity potential $|\phi(x,z)|$ when forced by $0.1$\,m waves which are incident from the left of the array of barriers. The frequencies of the incident waves shown here are the desired resonant frequencies, which were used to determine the parameters of the array. Specifically, these frequencies are (a) $\omega=1.6$\,s$^{-1}$, (b) $1.5$\,s$^{-1}$, (c) $1.4$\,s$^{-1}$, (d) $1.3$\,s$^{-1}$, (e) $1.2$\,s$^{-1}$, (f) $1.1$\,s$^{-1}$, (g) $1.0$\,s$^{-1}$, (h) $0.9$\,s$^{-1}$ (i) $0.8$\,s$^{-1}$. The position of the strongest local energy amplification moves towards the right as the frequency decreases.
  }
\label{fig:rainbow}
\end{figure}

Colour plots of $|\phi(x,z)|$ for the array parametrised by the submergences reported in table \ref{tab:sub_depth} are given in figure \ref{fig:rainbow} at all the frequencies $\omega^{(n)}$. As expected, the strongest local energy amplification in the graded array for a given frequency occurs in the cavity associated with that fundamental frequency. In results not shown here, this does not occur when waves are incident from the right. This is a consequence of the low transmission of high frequency waves discussed earlier.

In order to analyse this phenomenon further, we consider the propagation of water waves through an infinite periodic array of uniform vertical barriers. Using the present notation, this is the case where $x^{(n)}=nl$ and $d^{(n)}=d$ for all $n\in\Z$, where $l >0$ and $d\in(0,H)$ are fixed. Our intention is to locally approximate the wave field within a finite graded array of vertical barriers using solutions to the infinitely periodic case, which is reasonable provided that the grading is weak \citep{Bennetts2019}. Infinite periodic arrays of uniform vertical barriers can be studied using Bloch's theorem, which motivates seeking solutions of the form
\begin{equation}\label{bloch_thm}
    \phi(x+nl,z) = e^{\mathrm{i}qnl}\phi(x,z).
\end{equation}
In the above equation we have introduced the unknown Bloch wavenumber $q\in\C$. To compute the corresponding Bloch modes, \eqref{bloch_thm} is applied to the horizontal boundary of the unit cell $\{(x,z)|x\in(-l/2,l/2)\mbox{\ and\ }z\in(-H,0)\}$, which yields
\refstepcounter{equation}
$$
        \phi(l/2,z)=e^{\mathrm{i}q l}\phi(-l/2,z)\quad\mbox{and}\quad
    \frac{\partial \phi}{\partial x}(l/2,z)=e^{\mathrm{i}q l}\frac{\partial \phi}{\partial x}(-l/2,z).
     \eqno{(\theequation{\mathit{a},\mathit{b}})}
$$
These quasi-periodic boundary conditions are then solved in conjunction with (\ref{laplacian}--\ref{vanishing_barrier}), where suitable modifications are made so that $x\in(-l/2,l/2)$.

A numerical solution is obtained by solving a generalised eigenvalue problem, which incorporates the scattering matrix for a single vertical barrier \citetext{see e.g. \citealp{Peter2009}}. For each frequency $\omega$, propagating Bloch modes (i.e. $q\in\R$) are sought in the irreducible Brillouin zone $q\in [0,\pi/l]$. We obtain dispersion curves in the $q-\omega$ plane, which form so-called band diagrams. We remark that frequency intervals occupied by the dispersion curves are called passbands, and these frequencies are associated with a propagating mode which can travel through the infinite array without a change in amplitude. The slope of the dispersion curve $d\omega/dq$ determines the group velocity of this propagating mode. The frequency intervals in which there are no real-valued Bloch wavenumbers are called bandgaps. At these frequencies, waves cannot propagate without a change in amplitude.

\begin{figure}
  \centerline{\includegraphics[width=\textwidth]{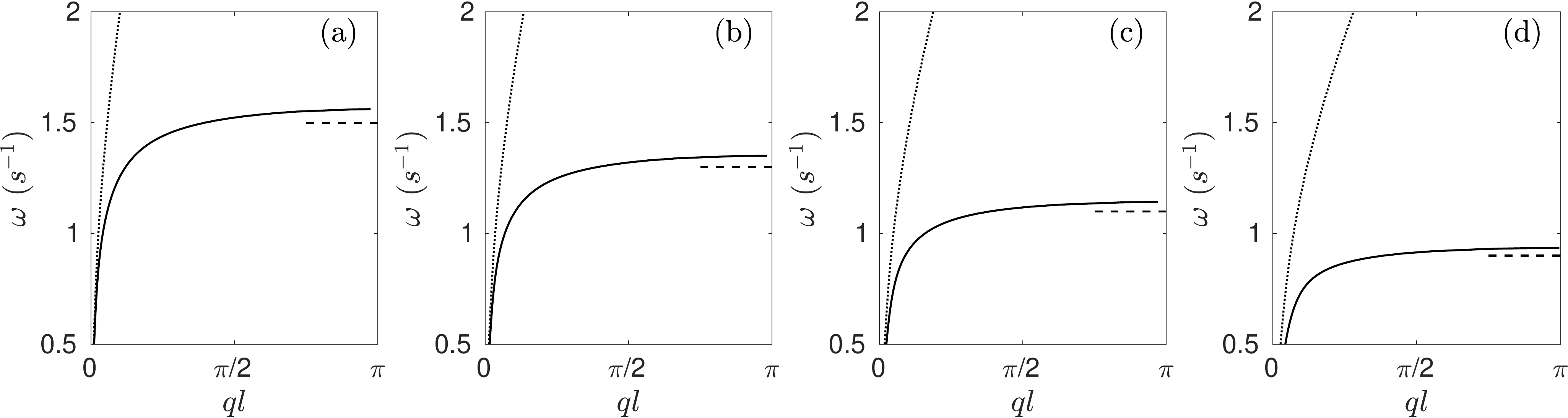}}
  \caption{(a-d) Dispersion curves for infinite periodic arrays of uniform vertical barriers (solid lines) in which $d=(d^{(n-1)}+d^{(n)})/2$ and $l=l^{(n)}$ for (a) $n=2$, (b) $4$, (c) $6$ and (d) $8$. Here, the values of $d^{(n)}$ and $l^{(n)}$ are those of the graded array shown in figure \ref{fig:rainbow}. In each subfigure, the dotted line is the dispersion curve for unobstructed waves. Notably, the bandgap begins above the fundamental frequency of a pair of barriers with separation $l$ and submergence $d$. These fundamental frequencies, which are represented by dashed lines, are (a) $\omega^{(n)}=1.5$\,s$^{-1}$, (b) $1.3$\,s$^{-1}$, (c) $1.1$\,s$^{-1}$ and (d) $0.9$\,s$^{-1}$.}
\label{fig:band_diagrams}
\end{figure}

Some examples of band diagrams with different barrier submergences and spacings are shown in figure \ref{fig:band_diagrams}. In each case, there is only one passband and one bandgap. We observe that the fundamental frequency of a pair of barriers with separation $l$ and submergence $d$ is a lower bound of the bandgap. The fundamental frequency decreases as the barriers become deeper, which coincides with the dispersion curves becoming more compressed. This mirrors a similar result found by \citet{Bennetts2018} for infinite two-dimensional arrays of C-shaped cylindrical resonators. In that setting, the resonant frequency of the isolated cylinder is an upper bound of the passband. In either cases, the bandgap can be shifted by modifying the resonant substructure of the array.

In the finite array of barriers, the potential surrounding each successive pair of barriers can be understood in terms of the corresponding Bloch problem. In particular, the dispersion curve of the infinite array of barriers with submergence $(d^{(n-1)}+d^{(n)})/2$ and spacing $l^{(n)}$ approximately determines the behavior in the $n$th cavity for a weakly graded structure, and in this setting it is referred to as the local dispersion curve. Since the fundamental frequencies of the cavities decrease from left to right, the local dispersion curves become more compressed and the frequency interval which supports propagating modes becomes narrower. Thus low frequency waves propagate further towards the right than high frequency waves.

Finite difference estimates of the group velocity of the propagating mode are tabulated in table \ref{tab:grp_vel} for all cavities and for a range of frequencies. Waves slow down as they propagate through the finite, graded array because the group velocity is a decreasing function of $n$, which attains a minimum before becoming undefined. This minimum typically occurs in the cavity whose fundamental frequency is the frequency of the wave. The wave can only attenuate beyond this point \citep[which is called the turning point][]{Romero-Garcia2013}, since propagation at this frequency is no longer supported by the array. A local energy amplification occurs near the turning point because the propagating mode has a relatively small group velocity, so its energy is localised for a relatively large duration. Moreover, the group velocity is a decreasing function of $\omega$ where it is defined, which means that higher frequency waves slow down more rapidly than lower frequency waves and amplify further towards the left, resulting in spatial separation. Thus, graded array of vertical barriers gives rise to the rainbow reflection effect.

\begin{table}
  \begin{center}
\def~{\hphantom{0}}
  \begin{tabular}{c|ccccccccc}
$\omega$ (s$^{-1}$) & $n=1$  & $n=2$  & $n=3$  & $n=4$  & $n=5$  & $n=6$  & $n=7$  & $n=8$  & $n=9$  \\ \hline
0.8& $4.48$ & $4.37$ & $3.85$ & $3.60$ & $2.98$ & $2.55$ & $1.79$ & $1.18$ & $0.42$ \\
0.9& $3.26$ & $3.14$ & $2.63$ & $2.38$ & $1.79$ & $1.41$ & $0.78$ & $0.36$ & ---    \\
1.0& $2.31$ & $2.20$ & $1.72$ & $1.50$ & $1.00$ & $0.69$ & $0.25$ & ---    & ---    \\
1.1& $1.64$ & $1.53$ & $1.10$ & $0.90$ & $0.49$ & $0.26$ & ---    & ---    & ---    \\
1.2& $1.13$ & $1.04$ & $0.65$ & $0.49$ & $0.17$ & $0.02$ & ---    & ---    & ---    \\
1.3& $0.74$ & $0.66$ & $0.33$ & $0.20$ & ---    & ---    & ---    & ---    & ---    \\
1.4& $0.44$ & $0.37$ & $0.12$ & $0.04$ & ---    & ---    & ---    & ---    & ---    \\
1.5& $0.22$ & $0.17$ & ---    & ---    & ---    & ---    & ---    & ---    & ---    \\
1.6& $0.09$ & $0.05$ & ---    & ---    & ---    & ---    & ---    & ---    & ---   
\end{tabular}
    \caption{Finite difference estimates of the group velocity (m s$^{-1}$) of waves travelling through the infinite array of barriers with $d=d^{(n-1)}$ and $l=l^{(n)}$, with frequency $\omega$. Dashes (---) are used to denote the absence of a propagating Bloch mode. The group velocity is a decreasing function of both $\omega$ and $n$.
    }
  \label{tab:grp_vel}
  \end{center}
\end{table}

However, the graded array does not give rise to the rainbow trapping effect. Because the unit cell of the Bloch problem possesses axial symmetry, the infinite array supports both left and right propagating modes at passband frequencies, which have equal and opposite group velocity. Consequently, the strong coupling between the right and left propagating modes near the turning point results in the wave changing direction \citep{He2012,Chaplain2020}. The left-travelling wave then accelerates as it propagates in reverse, carrying energy that is ultimately reflected by the array. In the following section, we will modify the system so that the localised energy can be absorbed within the array rather than being reflected.

\section{Solution of multiple piston cavity problem}\label{solution_multiple_pistons}
\subsection{Preliminaries}\label{bvp}
\begin{figure}
  \centerline{\includegraphics[width=\textwidth]{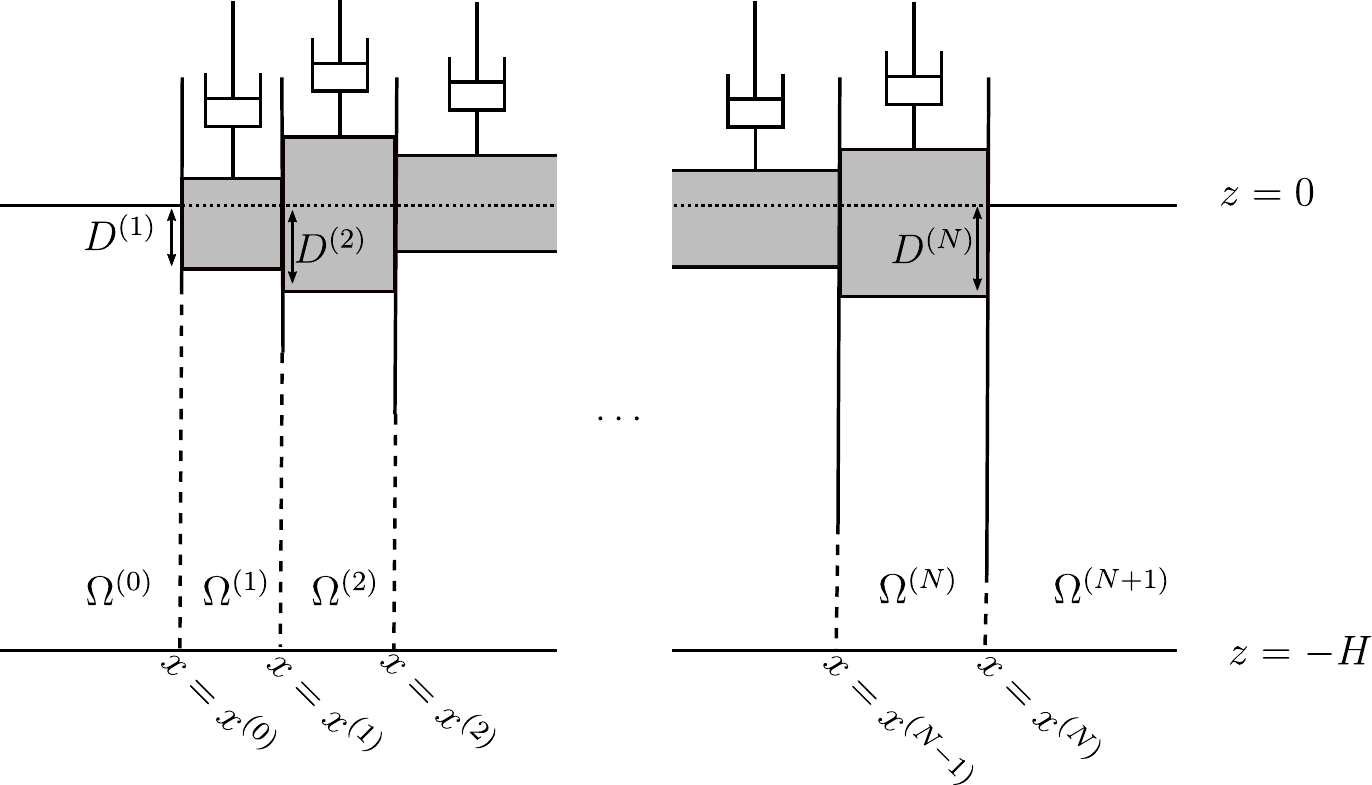}}
  \caption{Schematic of the multiple piston cavity problem. Viscous damping is symbolised by dashpots.}
\label{fig:schematic2}
\end{figure}

We adapt the problem considered in \textsection\ref{graded_arrays} by introducing $N$ rigid rectangular floating bodies, or `pistons', which completely occupy the horizontal gap between each adjacent pair of barriers. The pistons are restricted to move vertically in heave and are less dense than the fluid. Further, the density of each piston is $\rho_A^{(n)}/\delta^{(n)}$, where $\rho_A^{(n)}$ is the area density of the piston with respect to the wetted surface, and $\delta^{(n)}$ is the thickness. The motion of each piston is governed by a damping term proportional to the piston's vertical velocity, with the damping coefficient denoted $\mu^{(n)}$. A schematic of the geometry of this problem is given in figure \ref{fig:schematic2}.

We assume that pistons are undergoing small, time-harmonic oscillations about their equilibria. This implies that the submergence of the bottom face of each piston is
\begin{equation}
    z=\Real\left(s^{(n)}e^{-\mathrm{i}\omega t}\right)-D^{(n)},\label{eomsols}
\end{equation}
where $s^{(n)}$ and $D^{(n)}=\rho_A^{(n)}/\rho$ are the complex amplitude and the equilibrium submergence of the $n$\textsuperscript{th} piston respectively. Here, $\rho$ is the density of the fluid. The relevant forces which govern the dynamics of the pistons are due to hydrostatic pressure, hydrodynamic pressure, gravity and viscous damping. After some algebra, the linearised equations of motion can be stated as
\begin{equation}
    -\omega^2\rho_A^{(n)}l^{(n)}s^{(n)}-\mathrm{i}\omega\mu^{(n)} s^{(n)}+\rho g l^{(n)}s^{(n)}=\mathrm{i}\omega\rho \int_{x^{(n-1)}}^{x^{(n)}}\phi(x,-D^{(n)})dx,\label{eom2}
\end{equation}
where $l^{(n)}=x^{(n)}-x^{(n-1)}$ \citep{wehausen1971}.

We redefine the subregions $\Omega^{(n)}$ to be the sets of points beneath the $n$\textsuperscript{th} piston when at equilibrium, that is $\Omega^{(n)} = \{(x,z)|x\in(x^{(n-1)},x^{(n)})\text{ and }z\in(-H,-D^{(n)})\}$  for all $n\in\{1,\dots,N\}$. The definitions of $\Omega^{(0)}$ and $\Omega^{(N+1)}$ are unchanged. The velocity potential of the fluid $\phi$ satisfies the boundary value problem given by \eqref{laplacian}, \eqref{seabed}, \eqref{vanishing_barrier} and the additional conditions
\begin{subequations}
\begin{align}
    \partial_z\phi&=\frac{\omega^2}{g}\phi & \text{for } z=0\text{ and }x\in(-\infty,x^{(0)})\cup(x^{(n)},\infty)\\
    \partial_z\phi&=-\mathrm{i}\omega s^{(n)}&\mbox{\ where\ }x\in(x^{(n-1)},x^{(n)})\text{ and }z=-D^{(n)}. \label{kinematic_condition}
\end{align}
\end{subequations}
The potential is also subject to (\ref{limits}a,b), \eqref{corner_singularity}, \eqref{incident_field} and \eqref{sommerfeld}. 

\subsection{Derivation of a system of integral equations}
The expansion for the solution given in \eqref{separation_sol1} still applies in subregions $\Omega^{(0)}$ and $\Omega^{(N+1)}$, since there is no piston at the surface of the fluid. In the remaining subregions $\Omega^{(n)}$, we express the general solution as the sum of a homogeneous solution $\phi_h^{(n)}$ and a particular solution $\phi_p^{(n)}$, for all $n\in\{1,\dots,N\}$. The homogeneous solution satisfies \eqref{laplacian}, \eqref{seabed} and
\begin{equation}
    \partial_z\phi_h^{(n)}(x,-D^{(n)})=0,
\end{equation}
for $x\in(x^{(n-1)},x^{(n)})$. These three equations constitute a Sturm–Liouville problem in the $z$-direction. By using separation of variables and choosing an expansion centered at the midpoints of each piston $m^{(n)}=x^{(n-1)}+\tfrac{1}{2}l^{(n)}$ we obtain
\begin{multline}
    \phi_h^{(n)}(x,z)=A_0^{(n)}+B_0^{(n)}\left(x-m^{(n)}\right)+\sum_{m=1}^\infty \left(A_m^{(n)}\cosh\left(\kappa_m^{(n)} \left(x-m^{(n)}\right)\right)\right.\\
    \left.+B_m^{(n)}\sinh\left(\kappa_m^{(n)} \left(x-m^{(n)}\right)\right)\right)\frac{\psi_m^{(n)}(z)}{\cosh\left(\tfrac{1}{2}\kappa_m^{(n)}l^{(n)}\right)},\label{homogeneous_sol}
\end{multline}
where $\kappa_m^{(n)}=m\pi/h^{(n)}$ in which $h^{(n)}=H-D^{(n)}$, and the vertical eigenfunctions have been defined as 
\begin{equation*}
    \psi_m^{(n)}(z)=\sqrt{2}\cos(\kappa_m^{(n)}(z+H))
\end{equation*}
for all $m\in\N$. These eigenfunctions satisfy the following orthogonality relation
\begin{equation}\label{orthonormality2}
    \int_{-H}^{-D^{(n)}}\psi_m^{(n)}(\xi)\psi_p^{(n)}(\xi)d\xi = h^{(n)}\delta_{m p},
\end{equation}
for all $m,p\in\{0\}\cup\N$ and for all $n\in\{1,\dots,N\}$. 

The particular solution $\phi_p^{(n)}$ satisfies \eqref{laplacian}, \eqref{seabed} and \eqref{kinematic_condition}. We find a particular solution with the ansatz that $\phi_p^{(n)}$ is quadratic in both $x$ and $z$, which gives
\begin{equation}
    \phi_h^{(n)}(x,z)=-\frac{\mathrm{i}\omega s^{(n)}}{2h^{(n)}}\left((z+H)^2-\left(x-m^{(n)}\right)^2\right),\label{particular_sol}
\end{equation}
where we have imposed symmetry about the horizontal midpoints $x=m^{(n)}$. This treatment of the inhomogeneous boundary condition is in complete analogy to the solution obtained by \citet{Yeung1981} for the problem of an oscillating vertical circular cylinder in a fluid of finite depth.

In order to solve the multiple scattering problem for the unknown coefficients, we introduce auxiliary functions
\begin{equation}\label{auxiliary_def}
    u^{(n)}(z) = \left\{\begin{array}{ll}
    \lim_{x\to x_j}\partial_x\phi(x,z)&\mbox{for\ }z<-d^{(n)}\\
    0&\mbox{for\ }z\geq -d^{(n)}.
    \end{array}\right.
\end{equation}
Applying (\ref{limits}b) as $x \to x^{(0)-}$ and $x \to x^{(N)+}$ to \eqref{separation_sol1} yields
\begin{align}
    u^{(0)}(z)&=\mathrm{i}k_0A_0^{(0)}\psi_0^{(0)}(z)-\sum_{m=0}^\infty \mathrm{i}k_mB_m^{(0)}\psi_m^{(0)}(z)\\
   u^{(N)}(z)&=-\mathrm{i}k_0B_0^{(N+1)}\psi_0^{(0)}(z)+\sum_{m=0}^\infty \mathrm{i}k_mA_m^{(N+1)}\psi_m^{(0)}(z).
\end{align}
After referring to \eqref{orthonormality}, we obtain the following expressions for the scattered wave coefficients
\begin{align}
    B_m^{(0)}&=
    A_0^{(0)}\delta_{0m}+\frac{\mathrm{i}}{k_m H}\mathcal{P}^{(0)}(u^{(0)},\psi_m^{(0)})\label{b0}\\
    A_m^{(N+1)}&=
    B_0^{(N+1)}\delta_{0m}-\frac{\mathrm{i}}{k_m H}\mathcal{P}^{(N)}(u^{(N)},\psi_m^{(0)}) \label{aN}
\end{align}
where $\mathcal{P}^{(n)}(\cdot,\cdot)$ is the real inner product over $(-H,-d^{(n)})$. Similarly, applying (\ref{limits}b) as $x \to x^{(n-1)+}$ and $x \to x^{(n)-}$ gives, after referring to \eqref{homogeneous_sol} and \eqref{particular_sol} and invoking \eqref{orthonormality2},
\begin{align}
    A_m^{(n)}&=\frac{\coth\left(\tfrac{1}{2}\kappa_m^{(n)}l^{(n)}\right)}{2m\pi}\left(\mathcal{P}^{(n)}(u^{(n)},\psi_m^{(n)})-\mathcal{P}^{(n-1)}(u^{(n-1)},\psi_m^{(n)})\right)\label{Am}\\
    B_m^{(n)}&=\frac{1}{2m\pi}\left(\mathcal{P}^{(n)}(u^{(n)},\psi_m^{(n)})+\mathcal{P}^{(n-1)}(u^{(n-1)},\psi_m^{(n)})\right).\label{Bm}
\end{align}
Then, applying (\ref{limits}a) as $x\to x^{(n)}$ for all $n\in\{0,\dots,N\}$ and using \eqref{b0}, \eqref{aN}, \eqref{Am} and \eqref{Bm}, we obtain the following system of integral equations
\begin{multline}\label{int1}
    \int_{-H}^{-d^{(0)}}(K^{(0)}(z,\xi)+K^{(1)}(z,\xi))u^{(0)}(\xi)d\xi-\int_{-H}^{-d^{(1)}}L^{(1)}(z,\xi)u^{(1)}(\xi)d\xi\\
    +\frac{\mathrm{i}\omega s^{(1)}}{2h^{(1)}}\left((z+H)^2-\tfrac{1}{4}(l^{(1)})^2\right)-A_0^{(1)}  +\tfrac{1}{2}B_0^{(1)}l^{(1)}=-2A_0^{(0)}\psi_0^{(0)}(z),
\end{multline}
\begin{multline}\label{int2}
    -\int_{-H}^{-d^{(n-1)}}L^{(n)}(z,\xi)u^{(n-1)}(\xi)d\xi+\int_{-H}^{-d^{(n)}}(K^{(n)}(z,\xi)+K^{(n+1)}(z,\xi))u^{(n)}(\xi)d\xi\\
    -\int_{-H}^{-d^{(n+1)}}L^{(n+1)}(z,\xi)u^{(n+1)}(\xi)d\xi+\frac{\mathrm{i}\omega}{2}\left(\frac{s^{(n+1)}}{h^{(n+1)}}\left((z+H)^2-\tfrac{1}{4}(l^{(n+1)})^2\right)\right.\\ \left.-\frac{s^{(n)}}{h^{(n)}}\left((z+H)^2-\tfrac{1}{4}(l^{(n)})^2\right)\right)+A_0^{(n)}  +\tfrac{1}{2}B_0^{(n)}l^{(n)}-A_0^{(n+1)}  +\tfrac{1}{2}B_0^{(n+1)}l^{(n+1)}=0,
\end{multline}
\begin{multline}\label{int3}
    -\int_{-H}^{-d^{(N-1)}}L^{(N)}(z,\xi)u^{(N-1)}(\xi)d\xi+\int_{-H}^{-d^{(N)}}(K^{(N)}(z,\xi)+K^{(0)}(z,\xi))u^{(N)}(\xi)d\xi\\
    -\frac{\mathrm{i}\omega s^{(N)}}{2h^{(N)}}\left((z+H)^2-\tfrac{1}{4}(l^{(N)})^2\right)+A_0^{(N)} +\tfrac{1}{2}B_0^{(N)}l^{(N)}=2B_0^{(N+1)}\psi_0^{(0)}(z),
\end{multline}
where \eqref{int2} holds for all $n\in\{1,\dots,N-1\}$ and we have defined the integral kernels
\begin{align*}
    K^{(0)}(z,\xi)&=\sum_{m=0}^\infty \frac{\mathrm{i}}{k_mH}\psi_m^{(0)}(z)\psi_m^{(0)}(\xi),\\
    K^{(n)}(z,\xi)&=\sum_{m=1}^\infty \frac{\coth(\kappa_m^{(n)}l^{(n)})}{m\pi}\psi_m^{(n)}(z)\psi_m^{(n)}(\xi),\\
    L^{(n)}(z,\xi)&=\sum_{m=1}^\infty \frac{\csch(\kappa_m^{(n)}l^{(n)})}{m\pi}\psi_m^{(n)}(z)\psi_m^{(n)}(\xi).
\end{align*}

\subsection{Numerical solution using a Galerkin method}
To find an approximate solution to the system of integral equations, we first suppose that for each $n=0,\dots,N$, the auxiliary function $u^{(n)}$ can be expanded in terms of an appropriately chosen basis $\{v_j^{(n)}:(-H,-d^{(n)})\to\C|j\in\N\}$ over the interval $(-H,-d^{(n)})$. We then approximate each auxiliary function $u^{(n)}$ by
\begin{equation}\label{aux}
    u^{(n)}(z) \approx \begin{cases}
    \sum_{j=1}^{M_{\textrm{aux}}} c_j^{(n)}v_j^{(n)}(z)&\mbox{for\ }z<-d^{(n)}\\
    0&\mbox{for\ }z\geq -d^{(n)}.
    \end{cases}
\end{equation}
where we have truncated each expansion at some sufficiently large $M_{\textrm{aux}}\in\N$.

After substitution of \eqref{aux} into the system \eqref{int1}--\eqref{int3}, we multiply each expression by $v_p^{(n)}(z)$, for some $p\in\{1,\dots,M_{\textrm{aux}}\}$. Here, $n\in\{0,\dots,N\}$ is the index of the barrier associated with each integral equation. For instance, \eqref{int1} was derived by applying (\ref{limits}a) at $x^{(0)}$, so this expression is multiplied by $v_p^{(0)}(z)$ following substitution of \eqref{aux}. Integrating each of these expressions with respect to $z$ over $(-H,-d^{(n)})$ subsequently yields the following linear system of equations
\begin{multline}
    \sum_{m=0}^\infty\mathcal{U}_{mp}^{(0)}\mathcal{K}_{m}^{(0)}\sum_{j=1}^{M_{\textrm{aux}}} \mathcal{U}_{mj}^{(0)}c_j^{(0)}+\sum_{m=1}^\infty\mathcal{V}_{mp}^{(1)}\left(\mathcal{K}_{m}^{(1)}\sum_{j=1}^{M_{\textrm{aux}}}\mathcal{V}_{mj}^{(1)}c_j^{(0)}-\mathcal{L}_{m}^{(1)}\sum_{j=1}^{M_{\textrm{aux}}}\mathcal{U}_{mj}^{(1)}c_j^{(1)}\right)\\+\frac{\mathrm{i}\omega s^{(1)}}{2h^{(1)}}\left(\zeta_p^{(1)}-\tfrac{1}{4}(l^{(1)})^2\gamma_{p}^{(1)}\right)-A_0^{(1)}\gamma_{p}^{(1)} +\tfrac{1}{2}B_0^{(1)}l^{(1)}\gamma_{p}^{(1)}=-2\mathcal{U}_{0p}^{(0)}A_0^{(0)},\label{sys1}
\end{multline}
\begin{multline}
    -\sum_{m=1}^\infty \mathcal{U}_{mp}^{(n)}\left(\mathcal{L}_{m}^{(n)}\sum_{j=1}^{M_{\textrm{aux}}}\mathcal{V}_{mj}^{(n)}c_j^{(n-1)}-\mathcal{K}_{m}^{(n)}\sum_{j=1}^{M_{\textrm{aux}}}\mathcal{U}_{mj}^{(n)}c_j^{(n)}\right)\\
    +\sum_{m=1}^\infty \mathcal{V}_{mp}^{(n+1)}\left(\mathcal{K}_{m}^{(n+1)}\sum_{j=1}^{M_{\textrm{aux}}}\mathcal{V}_{mj}^{(n+1)}c_j^{(n)}-\mathcal{L}_{m}^{(n+1)}\sum_{j=1}^{M_{\textrm{aux}}}\mathcal{U}_{mj}^{(n+1)}c_j^{(n+1)}\right)\\
    +\frac{\mathrm{i}\omega}{2}\left(\frac{s^{(n+1)}}{h^{(n+1)}}\left(\zeta_p^{(n)}-\tfrac{1}{4}(l^{(n+1)})^2 \gamma_{p}^{(n)}\right)-\frac{s^{(n)}}{h^{(n)}}\left(\zeta_p^{(n)}-\tfrac{1}{4}(l^{(n)})^2 \gamma_{p}^{(n)}\right)\right)\\
    +A_0^{(n)} \gamma_{p}^{(n)} +\tfrac{1}{2}B_0^{(n)}l^{(n)} \gamma_{p}^{(n)}-A_0^{(n+1)}  \gamma_{p}^{(n)} +\tfrac{1}{2}B_0^{(n+1)}l^{(n+1)} \gamma_{p}^{(n)}=0,\label{sys2}
\end{multline}
\begin{multline}
    -\sum_{m=1}^\infty \mathcal{U}_{mp}^{(N)}\mathcal{L}_{m}^{(N)}\sum_{j=1}^{M_{\textrm{aux}}}\mathcal{V}_{mj}^{(N)}c_j^{(N-1)}+\sum_{m=1}^\infty \mathcal{U}_{mp}^{(N)}\mathcal{K}_{m}^{(N)}\sum_{j=1}^{M_{\textrm{aux}}}\mathcal{U}_{mj}^{(N)}c_j^{(N)}\\+\sum_{m=0}^\infty \mathcal{V}_{mp}^{(N+1)}\mathcal{K}_{m}^{(0)}\sum_{j=1}^{M_{\textrm{aux}}}\mathcal{V}_{mj}^{(N+1)}c_j^{(N)}-\frac{\mathrm{i}\omega s^{(N)}}{2h^{(N)}}\left( \zeta_{p}^{(N)}-\tfrac{1}{4}(l^{(N)})^2 \gamma_{p}^{(N)}\right)\\+A_0^{(N)} \gamma_{p}^{(N)} +\tfrac{1}{2}B_0^{(N)}l^{(N)} \gamma_{p}^{(N)}=2 B_0^{(N+1)}\mathcal{V}_{0p}^{(N+1)},\label{sys3}
\end{multline}
where \eqref{sys2} holds for all $n\in\{1,\dots,N-1\}$ and we have defined
\begin{equation*}
    \mathcal{K}_{m}^{(0)}=\frac{\mathrm{i}}{k_m H},\quad     \mathcal{K}_{m}^{(n)}=\frac{\coth(\kappa_m^{(n)}l^{(n)})}{m\pi},\quad \mathcal{L}_{m}^{(n)}=\frac{\csch(\kappa_m^{(n)}l^{(n)})}{m\pi}.
\end{equation*}
Additionally, we have defined
\begin{subequations}
\begin{align}
    \mathcal{U}^{(0)}_{mj}&=\mathcal{P}^{(0)}(\psi_m^{(0)},v_j^{(0)}), &\mathcal{V}^{(N+1)}_{mj}&=\mathcal{P}^{(N)}(\psi_m^{(0)},v_j^{(N)}),\label{FS_IPs}\\
    \mathcal{U}^{(n)}_{mj}&=\mathcal{P}^{(n)}(\psi_m^{(n)},v_j^{(n)})&\mathcal{V}^{(n)}_{mj}&=\mathcal{P}^{(n-1)}(\psi_m^{(n)},v_j^{(n-1)})\label{Piston_IPs}\\
    \gamma_{j}^{(n)}&=\int_{-H}^{-d^{(n)}}v_j^{(n)}(\xi)d\xi,&\zeta_j^{(n)}&=\int_{-H}^{-d^{(n)}}(\xi+H)^2v_j^{(n)}(\xi)d\xi.\label{additional_IPs}
\end{align}
\end{subequations}

The singularities at $(x^{(n)},-d^{(n)})$ motivate the following choice of basis functions
\begin{equation}
    v_j^{(n)}(\xi) = \frac{2(-1)^{j-1}}{\pi  \sqrt{(H-d^{(n)})^2-(H+\xi)^2}}T_{2(j-1)}\left(\frac{H+\xi}{H-d^{(n)}}\right),
\end{equation}
in which $T_n$ is the Chebyshev polynomial of the first kind of degree $n$. This basis was first introduced by \citet{Porter1995a} to solve the scattering of linear waves by a single surface-piercing barrier. By using equation (1.10.2) in \citet{Bateman1954}, the integrals in \eqref{FS_IPs} and \eqref{Piston_IPs} can be evaluated as
\begin{align*}
    \mathcal{U}_{mj}^{(0)}&=(-1)^{j-1}\beta_{m}^{-1/2} I_{2(j-1)}(k_{m}(H-d^{(0)})),\\
    \mathcal{V}_{mj}^{(N+1)}&=(-1)^{j-1}\beta_{m}^{-1/2} I_{2(j-1)}(k_{m}(H-d^{(N)})),\\
    \mathcal{U}_{mj}^{(n)}&=\sqrt{2}J_{2(j-1)}(\kappa_m^{(n)}(H-d^{(n)})),\\
    \mathcal{V}_{mj}^{(n)}&=\sqrt{2}J_{2(j-1)}(\kappa_m^{(n)}(H-d^{(n-1)})),
\end{align*}
where $J_n$ is the Bessel function of the first kind of order $n$, and $I_n$ is the corresponding modified Bessel function. Furthermore, by applying the symmetry and orthogonality of the even Chebyshev polynomials, the integrals in \eqref{additional_IPs} can be evaluated as
\begin{equation}
      \gamma_{j}^{(n)}=\delta_{1j}\quad\mbox{and}\quad\zeta_j^{(n)}=\frac{(H-d^{(n)})^2}{4}\begin{cases}
      2&j=1\\
      -1&j=2\\
      0&j\geq 3.
      \end{cases}
\end{equation}

Equations \eqref{sys1}, \eqref{sys2} and \eqref{sys3} form a linear system of $(N+1)M_{\textrm{aux}}$ equations with $(N+1)M_{\textrm{aux}}+3N$ unknowns. To obtain a further $2N$ equations, we once again apply (\ref{limits}b) as $x \to x^{(n-1)+}$ and $x \to x^{(n)-}$. However, in contrast to the derivation of \eqref{Am} and \eqref{Bm}, we use the fact that $\psi_m^{(n)}$ is orthogonal to the constant function $1$ for all $m\in\N$. After referring to \eqref{additional_IPs}, we have
\begin{align}
    \sum_{j=1}^{M_{\textrm{aux}}}\gamma_{j}^{(n-1)}c_j^{(n-1)}+\frac{\mathrm{i}\omega s^{(n)}l^{(n)}}{2}-B_0^{(n)}h^{(n)}&=0\label{bonus_eq_1}\\
    \sum_{j=1}^{M_{\textrm{aux}}}\gamma_{j}^{(n)}c_j^{(n)}-\frac{\mathrm{i}\omega s^{(n)}l^{(n)}}{2}-B_0^{(n)}h^{(n)}&=0,\label{bonus_eq_2}
\end{align}
for $n\in\{1,\dots,N\}$.

The remaining $N$ equations are derived from the equations of motion of the pistons \eqref{eom2}. The integral of the hydrodynamic pressure is evaluated by invoking symmetry about ${x=m^{(n)}}$, then expressed in terms of the coefficients of the auxiliary functions by referring to \eqref{Am}, \eqref{aux} and \eqref{Piston_IPs}
\begin{multline}
    \int_{x^{(n-1)}}^{x^{(n)}}\phi(x,-D^{(n)})dx=-\frac{\mathrm{i}\omega s^{(n)}}{2}h^{(n)}l^{(n)}+\frac{\mathrm{i}\omega s^{(n)}(l^{(n)})^3}{24h^{(n)}}+l^{(n)}A_0^{(n)}\\+\frac{\sqrt{2}h^{(n)}}{\pi^2} \sum_{m=1}^\infty\frac{(-1)^m}{m^2}\left(\sum_{j=1}^{M_{\textrm{aux}}}\mathcal{U}_{mj}^{(n)}c_j^{(n)}-\sum_{j=1}^{M_{\textrm{aux}}}\mathcal{V}_{mj}^{(n)}c_j^{(n-1)}\right).\label{eomint}
\end{multline}
After some rearrangement, substitution of \eqref{eomsols} and \eqref{eomint} into \eqref{eom2} yields
\begin{equation}
    \sigma^{(n)}s^{(n)}-\mathrm{i}\omega\rho l^{(n)}A_0^{(n)}-\frac{\mathrm{i}\omega\rho\sqrt{2}h^{(n)}}{\pi^2}\sum_{m=1}^\infty\frac{(-1)^m}{m^2}\left(\sum_{j=1}^{M_{\textrm{aux}}}\mathcal{U}_{mj}^{(n)}c_j^{(n)}-\sum_{j=1}^{M_{\textrm{aux}}}\mathcal{V}_{mj}^{(n)}c_j^{(n-1)}\right)=0,
\end{equation}
where we have defined
\begin{equation}
    \sigma^{(n)}=\rho_A^{(n)}l^{(n)}\omega^2+\rho g l^{(n)} -\frac{\omega^2\rho h^{(n)}l^{(n)}}{2}+\frac{\omega^2\rho(l^{(n)})^3}{24h^{(n)}}-\mathrm{i}\omega\mu^{(n)}.\label{eom3}
\end{equation}

Finally, we truncate the infinite sums in \eqref{sys1}, \eqref{sys2}, \eqref{sys3} after $M_{\textrm{ker}}\sim1000$ terms, then combine them with \eqref{bonus_eq_1}, \eqref{bonus_eq_2} and \eqref{eom3} to obtain a matrix equation of dimension $M_{\textrm{aux}}(N+1)+3N$. This allows us to simultaneously solve for the coefficients of the auxiliary functions $c_j^{(n)}$ for all $n\in\{0,\dots,N\}$ and $j\in\{1,\dots,M_{\textrm{aux}}\}$, as well as the quantities $A_0^{(n)}$, $B_0^{(n)}$ and $s^{(n)}$ for all $n\in\{1,\dots,N\}$. The size of $M_{\textrm{aux}}$ must be chosen in accordance with $M_{\textrm{ker}}\sim1000$. By choosing $M_{\textrm{ker}}=1000$ and $M_{\textrm{aux}}=50$, we obtain 4-digit accuracy in the unknown coefficients $R$, $T$ and $s^{(n)}$. The convergence properties of this class of Galerkin method is discussed in more detail by \citet{Linton2001}.

The unknown wave amplitudes in regions $\Omega^{(0)}$ and $\Omega^{(N+1)}$ can then be recovered by substituting \eqref{aux} into \eqref{b0} and \eqref{aN}. Referring to \eqref{FS_IPs} then gives
\begin{equation*}
    B_m^{(0)}=A_0^{(0)}\delta_{0m}+\mathcal{K}_{m}^{(0)}\sum_{j=1}^{M_{\textrm{aux}}}\mathcal{U}^{(0)}_{mj}c_j^{0}\quad\mbox{\ and\ }\quad
    A_m^{(N+1)}=B_0^{(N+1)}\delta_{0m}-\mathcal{K}_{m}^{(0)}\sum_{j=1}^{M_{\textrm{aux}}}\mathcal{V}^{(N+1)}_{mj}c_j^{N}.
\end{equation*}
Similarly, the remaining unknown coefficients can be recovered after additionally referring to \eqref{Am} and \eqref{Bm}
\begin{align*}
    A_m^{(n)}&=\frac{\coth\left(\kappa_m^{(n)}l^{(n)}/2\right)}{2m\pi}\left(\sum_{j=1}^{M_{\textrm{aux}}}\mathcal{U}_{mj}^{(n)}c_j^{(n)}-\sum_{j=1}^{M_{\textrm{aux}}}\mathcal{V}_{mj}^{(n)}c_j^{(n-1)}\right),\\
    B_m^{(n)}&=\frac{1}{2m\pi}\left(\sum_{j=1}^{M_{\textrm{aux}}}\mathcal{U}_{mj}^{(n)}c_j^{(n)}+\sum_{j=1}^{M_{\textrm{aux}}}\mathcal{V}_{mj}^{(n)}c_j^{(n-1)}\right).
\end{align*}

\section{Energy absorption}\label{Energy_absorption_sec}
Due to the presence of damping, the class of devices introduced in \textsection\ref{solution_multiple_pistons} is capable of absorbing energy from the incident waves. Here, absorption is defined as $\alpha(\omega)= 1-|T|^2-|R|^2$, where the transmission and reflection coefficients $T$ and $R$ are calculated when the system is forced from the left by a plane wave of frequency $\omega$. By considering the energy flux of the incident and scattered plane waves \citep{Linton2001} and the time-average work done by the damping force, we can write
\begin{equation}\label{piston_contribution}
    \alpha(\omega)=\sum_{n=1}^N \alpha^{(n)}(\omega),\quad\textrm{where}\quad\alpha^{(n)}(\omega)=\frac{\omega\mu^{(n)}}{\rho k_0 H}\left|\frac{s^{(n)}}{A_0^{(0)}}\right|^2.
\end{equation}
This second form allows us to consider the contribution of each piston to the absorption by the entire device. It is clear that the minimum value of absorption is $\alpha=0$, which occurs, for example, in the absence of any damping. Further, the maximum value of absorption is $\alpha=1$, which occurs when no energy is scattered by the device (i.e. $R=T=0$).

\subsection{Absorption by a single cavity}\label{SCA_sec}
We first consider absorption by a single cavity absorber (SCA), here defined as a single piston surrounded by a pair of barriers. When the two barriers are identical, an established theoretical result implies that the maximum value of absorption is $\alpha=0.5$ \citep{Budal1975,Evans1976,Newman1977,Mei1976}. For any given frequency, this maximum can be attained by tuning the damping coefficient of the piston and the submergence of the barriers. This tuning is simplified by fixing $l^{(1)}=d^{(0)}/5$ and $\rho_A^{(1)}=500$\,kg\,m$^{-2}$. The parameter which maximise absorption for the frequencies considered in \textsection\ref{fund_res_sec} are presented in table \ref{tab:mu_opt}, where we use a tilde to distinguish parameters that relate to SCAs. We observe that the submergence depths are nearly identical to those in table \ref{tab:sub_depth}, which were obtained by maximising the time-average power of the free surface oscillations. This is because the fundamental resonance considered in \textsection\ref{fund_res_sec} is essentially a vertical fluid motion between the barriers, which is the fluid motion that maximises energy absorption by a heaving piston.

\begin{table}
  \begin{center}
\def~{\hphantom{0}}
  \begin{tabular}{cccccccccc}
$n$ & 1    & 2    & 3    & 4    & 5    & 6    & 7    & 8     & 9     \\
$\omega^{(n)}$ (s$^{-1})$& 1.6  & 1.5  & 1.4  & 1.3  & 1.2  & 1.1  & 1.0  & 0.9   & 0.8   \\
$\tilde{d}^{(n)}$ (m) & 3.41 & 3.88 & 4.46 & 5.17 & 6.06 & 7.22 & 8.74 & 10.78 & 13.58 \\
$\tilde{\mu}^{(n)}$ (kg m$^{-1}$ s$^{-1}$) & 107  & 130  & 159  & 199  & 253  & 332  & 460  & 692   & 1170 
\end{tabular}
\caption{The optimised parameters of a SCA which maximise absorption at $\omega^{(n)}$, i.e. $\alpha(\omega^{(n)})=0.5$ is attained. }
  \label{tab:mu_opt}
  \end{center}
\end{table}

We remark that larger values of $\alpha$ can be attained when $d^{(1)}>d^{(0)}$. Perfect absorption (i.e. $\alpha=1$) occurs when the wave radiated by the piston destructively interferes with the scattered wave, which is impossible for a symmetric device oscillating in one mode of motion \citetext{see \citealp{Falnes1998} for more details}. However, in the limiting case where $d^{(1)}=H$, the scattered wave propagates in one direction only as no transmission is allowed. This implies that complete destructive interference of the scattered wave by the radiated wave produced by a heave-restricted piston is possible. This limiting case is analogous to the oscillating water columns positioned against a wall considered by \citet{Evans1995}, where perfect absorption at a given frequency was achieved. 

\subsection{Optimised absorption at $N$ frequencies}\label{Jimenez_absorption}
Consider now an array of absorbing cavities whose absorption spectrum has maxima at the $N=9$ prescribed frequencies considered in \textsection\ref{SCA_sec}. We construct this array by modifying the algorithm which was used by \citet{Jimenez2017} to design an acoustic rainbow absorber. The algorithm is a sequence of optimisation problems, whereby the size of the device and the set of frequencies at which absorption is maximised both increase with each iteration. The details of the algorithm are provided in the following paragraph.

For $n$ iteratively incrementing from 1 to $N$, we maximise $\sum_{j=1}^{n} \alpha(\omega^{(N-j+1)})$ for a device consisting of $n$ cavities, by tuning $d_n^{(j)}$ for $j\in\{0,\dots,n\}$, and $\mu_n^{(j)}$ for $j\in\{1,\dots,n\}$. Here, the subscript $n$ refers to the stage of the iteration. The optimisation is performed while fixing ${l_n^{(j)}=(d_n^{(j-1)}+d_n^{(j)})/10}$ for $j\in\{1,\dots,n\}$, and with the constraints ${\tilde{d}^{(N-n)}<d_n^{(0)}<\dots<d_n^{(n)}<0.8H}$. When $n=N$, these constraints instead become ${2D^{(0)}<d_N^{(0)}<\dots<d_N^{(N)}<0.8H}$. For $n=1$ the initial guesses of the parameters are $d_n^{(0)}=d_n^{(1)}=\tilde{d}^{(N)}$ and $\mu_n^{(1)}=\tilde{\mu}^{(N)}$. At each subsequent stage $n>1$, we initialise the optimisation with ${d_n^{(0)}=\tilde{d}^{(N-n+1)}}$, ${d_n^{(j)}=d_{n-1}^{j-1}}$ for $j\in\{1,\dots,n\}$, $\mu_n^{(1)}=\tilde{\mu}^{(N-n+1)}$ and $\mu_n^{(j)}=\mu_{n-1}^{(j-1)}$ for $j\in\{2,\dots,n\}$. In other words, a new cavity which maximises absorption near $\omega^{(N-n+1)}$ is added to the left of the array constructed during step $n-1$.

\begin{figure}
    \centering
    \includegraphics[width=\textwidth]{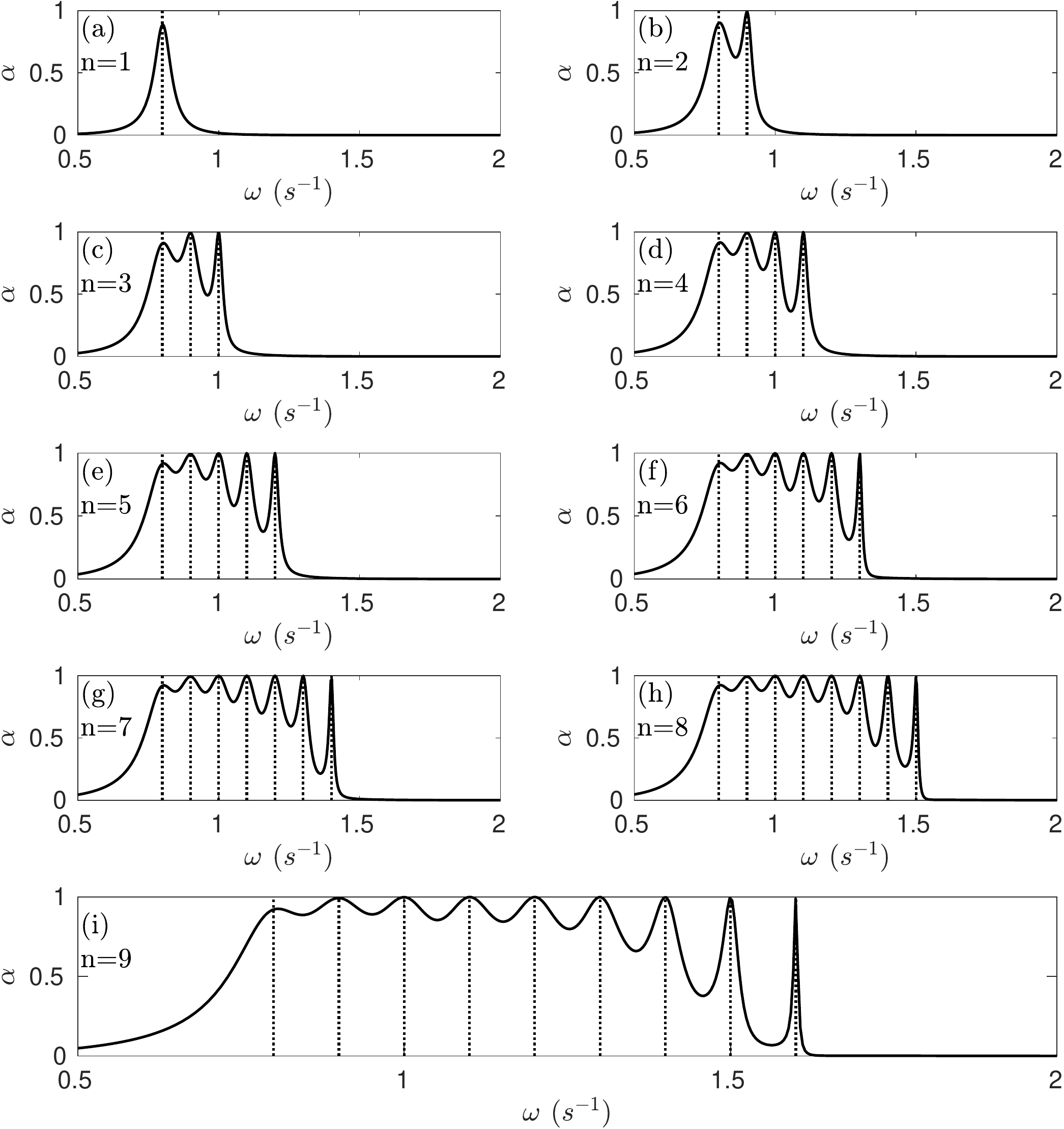}
    \caption{Absorption spectra of the devices obtained using the steps outlined in \textsection\ref{Jimenez_absorption}, which is a modification of that used by \citet{Jimenez2017}. The frequencies at which absorption is maximised at each stage of the algorithm are denoted by vertical dotted lines. As desired, the peaks of the spectra coincide with the desired peaks of absorption.}
    \label{fig:Jimenez_absorption}
\end{figure}

The absorption spectra of all of the intermediate devices obtained by this process are given in figure \ref{fig:Jimenez_absorption}, and the parameters of the final device with $n=9$ cavities are given in table \ref{tab:Jimenez_absorber}. As desired, we find that the peaks of the spectra of all intermediate devices coincide with the prescribed frequencies. In analysis not presented here, we find that at the $j$th peak of each spectrum, the largest contribution to absorption is by the $(n-j+1)$th piston. For the final device, we obtain $\alpha(\omega^{(j)})>0.98$ for $j\in\{1,\dots,N-1\}$. The constraint that $d^{(N)}<0.8H$ (which was imposed to ensure the solution converges rapidly) hinders absorption at the lowest frequency, where we obtain $\alpha(\omega^{(N)})\approx0.93$. In contrast, \citet{Jimenez2017} was able to obtain an acoustic device for which $\alpha=1$ at all $N$ frequencies as the result of using a less-restrictive absorption mechanism. We also observe that the peaks of the spectrum in figure \ref{fig:Jimenez_absorption}i become increasingly sharp as $\omega$ increases. As a consequence, absorption becomes increasingly lower in the intervals between the prescribed frequencies. Our next task is to rectify this by using a broadband absorption metric to design a rainbow absorber (RA), whose absorption spectrum remains close to $1$ over the entire interval $[\omega^{(N)},\omega^{(1)}]$.

\begin{table}
      \begin{center}
\def~{\hphantom{0}}
    \begin{tabular}{ccccccccccc}
$j$         & 0    & 1    & 2    & 3    & 4    & 5    & 6    & 7    & 8     & 9     \\
$d_N^{(j)}$ (m)  & 3.54 & 3.57 & 4.02 & 4.61 & 5.39 & 6.42 & 7.76 & 9.51 & 12.26 & 16.00 \\
$l_N^{(j)}$ (m)  & $-$  & 0.71 & 0.76 & 0.86 & 1.00 & 1.18 & 1.42 & 1.73 & 2.18  & 2.83  \\
$\mu_N^{(j)}$ (kg m$^{-1}$ s$^{-1}$) &  $-$    & 1    & 28   & 112  & 246  & 437  & 660  & 950  & 1767  & 3318 
\end{tabular}
    \caption{Parameters of the device obtained using the modified algorithm of \citet{Jimenez2017}. The absorption spectrum of this device is shown in figure \ref{fig:Jimenez_absorption}i.}
    \label{tab:Jimenez_absorber}
    \end{center}
\end{table}

\subsection{Optimised broadband absorption}
Using a similar metric to \citet{Jimenez2017}, we quantify the broadband performance of the RA as the average value of $\alpha$ over the interval $[\omega^{(N)},\omega^{(1)}]$, i.e.
\begin{equation}
    E=\frac{1}{\omega^{(1)}-\omega^{(N)}}\int_{\omega^{(N)}}^{\omega^{(1)}}\alpha(\omega)d\omega.
\end{equation}
It is clear that $E\in[0,1]$. Naively, we can increase $E$ by increasing the length of the cavities, since this decreases the $Q$-factors of the resonant peaks and allows them to overlap more. Broadband performance can also be increased by adding more appropriately-tuned cavities to the RA (i.e. increasing $N$), since this increases the number of absorption peaks in a given interval. Both of these avenues to increasing $E$ also increase the total length of the RA given by $L=\sum_{n=1}^N l^{(n)}$, which is undesirable due to practical considerations. This motivates the present problem of maximising $E$ while constraining $L$, where we will fix $N=9$.

To achieve this, we simultaneously tune the parameters $d^{(n)}$, $\mu^{(n)}$ and $l^{(n)}$ in order to maximise $E$, subject to the constraint that $L<\lambda_{0.8}/5$, where $\lambda_{0.8}\approx 86.4$\,m is the wavelength of a plane wave with frequency $\omega^{(N)}=0.8$\,s$^{-1}$. We also require that $l^{(n)}$ and $\mu^{(n)}$ are increasing, and that $2D^{(1)}<d^{(0)}<\dots<d^{(N)}<0.8H$. The initial guess is the device with $9$ cavities obtained in \textsection\ref{Jimenez_absorption}, which has an absorption peak at $9$ discrete frequencies and for which $E\approx0.787$. Our optimisation is conducted using the MATLAB routine \textit{fmincon}, and $E$ is approximated using trapezoidal quadrature.

\begin{figure}
    \centering
    \includegraphics[width=\textwidth]{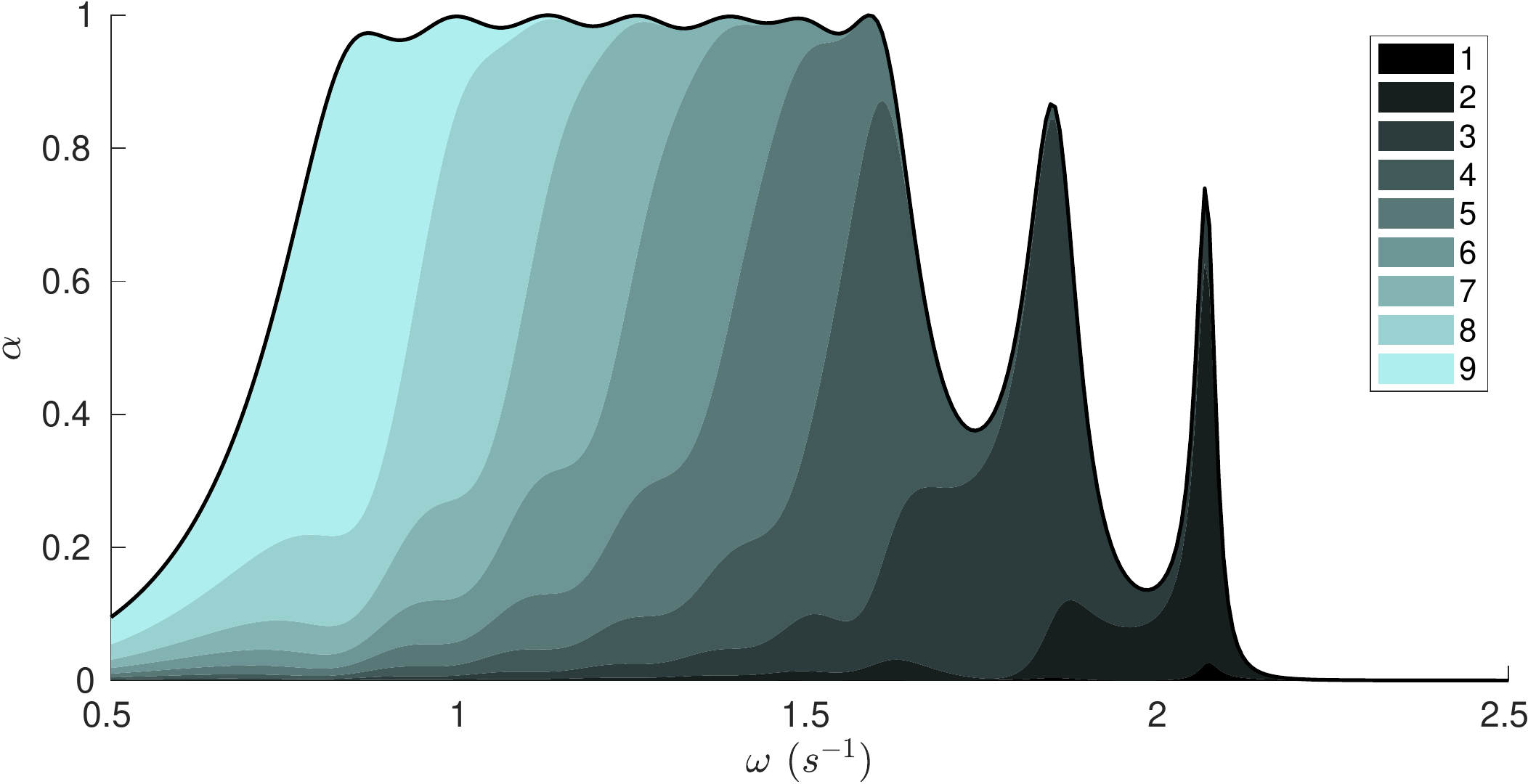}
    \caption{(Solid line) the absorption spectrum of the optimised RA. The coloured areas under the curve show the relative contribution of each piston to the total absorption by the device.}
    \label{fig:optimised_absorption}
\end{figure}

\begin{figure}
    \centering
    \includegraphics[width=\textwidth]{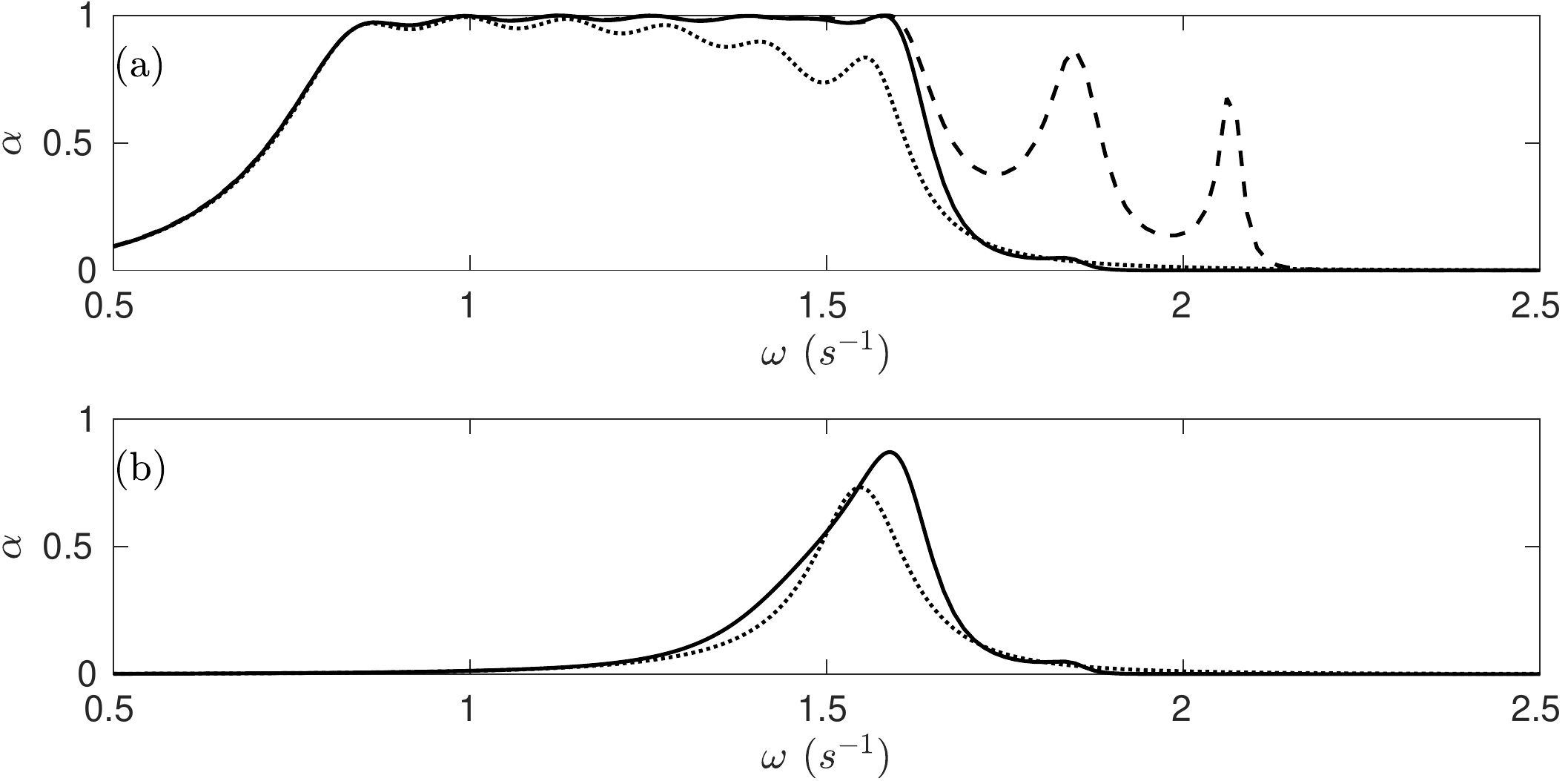}
    \caption{(a) The absorption spectrum of the optimised RA where the first three cavities have been removed (dotted line) and where the damping coefficients of the first three pistons are changed to zero (solid line). The absorption spectrum of the unaltered RA also shown in figure \ref{fig:optimised_absorption} is included for comparison (dashed line). (b) The absorption spectra of devices (i) (dotted line) and (ii) (solid line) described in the text. As was the case in (a), device (ii) differs from device (i) by the inclusion of 3 non-absorbing cavities. 
    }
    \label{fig:coupling}
\end{figure}

The parameters of the optimised RA, which attains a value of $E\approx0.982$, are given in table \ref{tab:Broadband_RA_parameters}, and its absorption spectrum is shown in figure \ref{fig:optimised_absorption}. The spectrum is predictably flat over the interval $[0.8,1.6]$\,s$^{-1}$, and we observe two additional absorption peaks at $\omega\approx1.84$\,s$^{-1}$ and $\omega\approx2.07$\,s$^{-1}$. By using \eqref{piston_contribution} to partition the area under the absorption spectrum in figure \ref{fig:optimised_absorption}, we find that the two additional peaks are the resonances of the third and second cavity respectively. Additionally, the first three pistons have a minor contribution to absorption over the target interval, which can be demonstrated by computing
\begin{equation*}
    \frac{1}{\omega^{(1)}-\omega^{(N)}}\sum_{n=1}^3\int_{\omega^{(N)}}^{\omega^{(1)}}\alpha^{(n)}(\omega)d\omega<0.035.
\end{equation*}
However, the RA obtained by removing the first three cavities, with absorption spectrum given by the dotted line in figure \ref{fig:coupling}a, only attains a value of $E\approx0.909$. Therefore the purpose of first three cavities cannot be understood in terms of absorption alone.

\begin{table}
      \begin{center}
\def~{\hphantom{0}}
    \begin{tabular}{ccccccccccc}
$n$& 0    & 1    & 2    & 3    & 4    & 5    & 6    & 7    & 8    & 9     \\
$d^{(n)}$ (m)& 1.80 & 1.80 & 2.18 & 3.13 & 3.63 & 4.35 & 5.47 & 7.07 & 9.91 & 16.00 \\
$l^{(n)}$ (m)& $-$  & 1.47 & 1.47 & 1.47 & 1.47 & 1.47 & 1.50 & 1.84 & 2.56 & 3.55  \\
$\mu^{(n)}$ (kg m$^{-1}$ s$^{-1}$) & $-$  & 2    & 56   & 223  & 492  & 874  & 1320 & 1900 & 3533 & 6636 
\end{tabular}
    \caption{Parameters of the device designed by maximising the broadband performance metric $E$. The absorption spectrum of this device is shown in figure \ref{fig:optimised_absorption}.}
    \label{tab:Broadband_RA_parameters}
    \end{center}
\end{table}

Indeed, the primary role of cavities $1$--$3$ is to intensify the local energy amplification in cavities $4$--$9$, which results in greater absorption by the pistons in these cavities. We interpret that this is because the coupling between the incident plane wave and the resonant mode in the cavity where the amplification occurs is strengthened by the inclusion of cavities $1$--$3$. To illustrate this, we consider (i) a SCA with the same parameters as cavity $4$ of the optimised RA, and (ii) an array of $4$ cavities consisting of the first four cavities of the optimised RA, but where $\mu^{(n)}=0$ for $n\in\{1,2,3\}$. Hence both devices contain a single absorbing cavity whose parameters are identical, but device (ii) additionally contains 3 non-absorbing cavities positioned to the left. Cavities $5$--$9$ have been omitted from both devices in order to simplify their absorption spectra, which are contrasted in figure \ref{fig:coupling}b. Importantly, the absorption peak of device (ii) is significantly higher than that of device (i), which suggests that the resonant mode in the absorbing cavity of device (ii) is more strongly coupled to the incident plane wave. This coupling occurs sequentially via the propagating Bloch modes of the three non-absorbing cavities, through which the group velocity gradually decreases (recall from \textsection\ref{local_energy_amps_sec} that the group velocity of the resonant mode is close to zero). In device (i), the coupling between the plane wave and the resonant mode must occur directly, which evidently results in a less-intense energy amplification. In figure \ref{fig:coupling}a, this experiment is also conducted without omitting cavities $5$--$9$ (see solid line), so that both of the devices being compared have $6$ identical absorbing cavities. We observe that absorption over $[0.8,1.6]$\,s$^{-1}$ is reduced much less by setting the damping coefficient of the first three pistons to zero than by removing the first three cavities. This further demonstrates the important role that cavities $1$--$3$ have in facilitating the transfer of energy from the incident plane wave to pistons $4$--$9$ in the optimised RA. We conclude that the two additional absorption peaks in figure \ref{fig:optimised_absorption} are a by-product of the geometry that cavities $2$ and $3$ must have in order to optimise this energy transfer.

\section{Concluding remarks}
Graded arrays of surface piercing vertical barriers have been proposed as a model for studying the rainbow reflection of linear water waves. An example consisting of 10 barriers was studied, where the parameters were selected so that the fundamental frequency of each successive pair of barriers gradually decreases throughout the array. An incident plane wave causes a local energy amplification in this array in the pair of barriers with that fundamental frequency. Using band diagram calculations for the cognate infinite array problem, we demonstrated that the local energy amplification originates from the rainbow reflection effect, since broadband wave signals slow down and become spatially separated based on frequency.

The graded array was modified to incorporate absorption by adding damped, heave-restricted pistons between each adjacent pair of barriers. We solved the resulting boundary value problem using an integral equation/Galerkin technique, giving accurate numerical results. We found parameters for devices which maximise absorption over a discrete set of frequencies by modifying the algorithm of \citet{Jimenez2017}. Further tuning using a broadband absorption metric produced a device that achieves near-perfect absorption over a prescribed frequency interval. While several of the cavities in the optimised rainbow absorber absorb a negligable proportion of energy over this interval, they were shown to be indirectly important for transferring energy from the incident wave to other absorbing cavities.

The study of rainbow absorption presented here is a proof of concept that should motivate further exploration of the applications of rainbow phenomena to water waves. For instance, a more realistic three dimensional fluid domain containing a two-dimensional arrays of resonant absorbers could be considered. Future work should also seek to validate the local energy amplification predicted using linear water wave theory, which could involve the use of computational fluid dynamics software or wave-tank experiments. Finally, more realistic energy absorption models, including models of wave-to-wire energy conversion, could be explored in the context of rainbow absorption for applications to the design of wave energy converters.

\section{Declaration of Interests}
The authors report no conflict of interest.

\appendix
\section{}\label{appA}
This appendix contains a discussion of how the solution to the underdetermined system of equations given in \eqref{underdetermined_sys} is uniquely specified in the paper. This system can be written in matrix form as
\begin{equation}\label{matrix_eq}
    \mathsfbi{C}\boldsymbol{d}=\boldsymbol{\tilde{d}},
\end{equation}
where $\boldsymbol{d}\in\R^{N+1}$ has entries $d^{(n+1)}$ and $\boldsymbol{\tilde{d}}\in\R^N$ has entries $\tilde{d}^{(n)}$. The $N\times(N+1)$ matrix $\mathsfbi{C}$ is bidiagonal, where each of the diagonal and superdiagonal entries are $0.5$. The null space of $\mathsfbi{C}$ is the span of a single vector $\boldsymbol{v}$, where $(\boldsymbol{v})_j=(-1)^{(j-1)}$. The set of solutions to \eqref{matrix_eq} is the line $\{\mathsfbi{C}^+\boldsymbol{\tilde{d}}+t\boldsymbol{v}|t\in\R\}$, where $\mathsfbi{C}^+$ denotes the Moore-Penrose inverse of $\mathsfbi{C}$.

The set of points satisfying the constraints $0<d^{(0)}<d^{(1)}<\dots<d^{(N)}<H$ is the intersection of $N+1$ half-spaces, which is a bounded convex set. The set of feasible solutions to \eqref{matrix_eq} is the line segment $\mathcal{F}=\{\mathsfbi{C}^+\boldsymbol{\tilde{d}}+t\boldsymbol{v}|t\in[t_{\textrm{min}},t_{\textrm{max}}]\}$, where $t_{\textrm{min}},t_{\textrm{max}}\in\R$ are determined by the constraints. We select $\boldsymbol{d}\in\mathcal{F}$ which minimises $\sum_{n=1}^N(d^{(n)}-d^{(n-1)})^2$, so that the grading of the array is as weak as possible.

\bibliographystyle{unsrtnat}
\bibliography{library}

\end{document}